\documentclass[aps,floatfix,nofootinbib,amsmath,amssymb,amsfonts,notitlepage,superscriptaddress, twocolumn]{revtex4-2}

\usepackage[T1]{fontenc}
\usepackage[latin9]{inputenc}
\usepackage{lmodern} 
\usepackage{color}
\usepackage{bbold}
\usepackage{dsfont}
\usepackage{graphicx}
\usepackage{comment}
\usepackage[caption=false]{subfig}
\usepackage[colorlinks]{hyperref}
\usepackage[bottom]{footmisc}
\usepackage{enumerate}
\usepackage{float}
\usepackage{mathtools}

\usepackage{empheq}
\usepackage{tikz}
\usepackage[normalem]{ulem}
\DeclareMathAlphabet\mbc{OMS}{cmsy}{b}{n}
\usepackage{balance}

\begin{document}

\newcommand{\ks}[1]{{\textcolor{teal}{[KS: #1]}}}
\newcommand{\cj}[1]{{\textcolor{blue}{CJ: #1}}}

\global\long\def\eqn#1{\begin{align}#1\end{align}}
\global\long\def\vec#1{\overrightarrow{#1}}
\global\long\def\ket#1{\left|#1\right\rangle }
\global\long\def\bra#1{\left\langle #1\right|}
\global\long\def\bkt#1{\left(#1\right)}
\global\long\def\sbkt#1{\left[#1\right]}
\global\long\def\cbkt#1{\left\{#1\right\}}
\global\long\def\abs#1{\left\vert#1\right\vert}
\global\long\def\cev#1{\overleftarrow{#1}}
\global\long\def\der#1#2{\frac{{d}#1}{{d}#2}}
\global\long\def\pard#1#2{\frac{{\partial}#1}{{\partial}#2}}
\global\long\def\re{\mathrm{Re}}
\global\long\def\im{\mathrm{Im}}
\global\long\def\dd{\mathrm{d}}
\global\long\def\ddd{\mathcal{D}}
\global\long\def\hmb#1{\hat{\mathbf #1}}
\global\long\def\avg#1{\left\langle #1 \right\rangle}
\global\long\def\mr#1{\mathrm{#1}}
\global\long\def\mb#1{{\mathbf #1}}
\global\long\def\mc#1{\mathcal{#1}}
\global\long\def\tr{\mathrm{Tr}}
\global\long\def\dbar#1{\Bar{\Bar{#1}}}

\global\long\def\nth{$n^{\mathrm{th}}$\,}
\global\long\def\mth{$m^{\mathrm{th}}$\,}
\global\long\def\non{\nonumber}

\newcommand{\orange}[1]{{\color{orange} {#1}}}
\newcommand{\cyan}[1]{{\color{cyan} {#1}}}
\newcommand{\blue}[1]{{\color{blue} {#1}}}
\newcommand{\yellow}[1]{{\color{yellow} {#1}}}
\newcommand{\green}[1]{{\color{green} {#1}}}
\newcommand{\red}[1]{{\color{red} {#1}}}
\global\long\def\todo#1{\orange{{$\bigstar$ \cyan{\bf\sc #1}}$\bigstar$} }

\title{Decoherence and Brownian motion of a polarizable particle near a surface}

\author{Clemens Jakubec }
\email{clemens.jakubec@univie.ac.at}
\affiliation{Vienna Center for Quantum Science and Technology, Faculty of Physics, University of Vienna, Boltzmanngasse 5, A-1090 Vienna, Austria}
\affiliation{Wyant College of Optical Sciences, University of Arizona, Tucson, Arizona 85721, USA}

\author{Christopher Jarzynski}
\email{cjarzyns@umd.edu}
\affiliation{Department of Chemistry and Biochemistry, Institute for Physical Science and Technology,
Department of Physics, University of Maryland, College Park, Maryland 20742, USA}

\author{Kanu Sinha}
\email{kanu@arizona.edu}
\affiliation{Wyant College of Optical Sciences, University of Arizona, Tucson, Arizona 85721, USA}
\affiliation{Department of Physics, University of Arizona, Tucson, Arizona 85721, USA}

\begin{abstract}
    We analyze the classical and quantized center-of-mass motion of a polarizable particle interacting with the fluctuations of the electromagnetic (EM) field in the presence of a medium. As a polarizable particle is immersed in a thermal environment, the momentum impulses imparted by the field fluctuations lead to momentum diffusion and drag for the particle's \textit{classical} center of mass. When considering the \textit{quantized} center-of-mass motion of the particle, these very fluctuations gain information about its position, leading to decoherence in the position basis.
     We derive a position localization master equation for the particle's quantized center of mass, and examine its classical center-of-mass momentum diffusion, elucidating correspondences between classical and quantum Brownian motion of polarizable particles near media. 
\end{abstract}

\maketitle
\section{Introduction}

The noise of the quantized electromagnetic (EM) field engenders phenomena over a wide range of length scales, from atomic Lamb shift \cite{Lamb47} and spontaneous emission \cite{Milonni75} to attractive forces in a micromechanical device~\cite{Stange2022}.
Just as a pollen grain in the noisy environment of fluid molecules exhibits random walk, a polarizable particle interacting with the fluctuations of the EM field experiences Brownian motion~\cite{EinsteinBrown, EinsteinHopf}.

\begin{figure*} [t]\centering
    \includegraphics[width = \textwidth]
{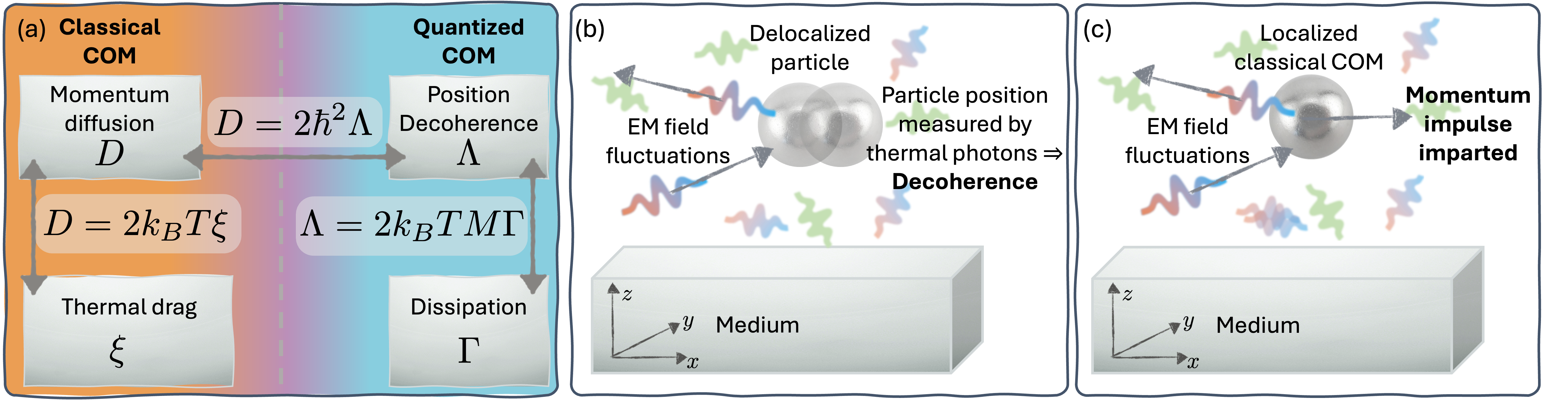}
    \caption{(a) Connections between Brownian motion of the classical and quantized center-of-mass motion of a polarizable point-particle immersed in a thermal EM environment. For the classical center of mass, momentum diffusion $D$ and thermal drag coefficient $\xi $ \cite{EinsteinHopf, Mkrtchian2003}  are related by the fluctuation-dissipation relation (FDR) $(D = 2 k_B T \xi )$. Similarly, the decoherence rate $(\Lambda)$ and dissipation $(\Gamma)$ for the quantized center-of-mass are connected via the FDR  $(\Lambda = 2 k_B T M \Gamma )$~\cite{BPBook}. Additionally, the center-of-mass decoherence rate $\Lambda$ is related to the momentum diffusion $ D$ via the relation $ D = 2 \hbar^2 \Lambda$~\cite{PWM_KS22}.  Schematic of a polarizable particle interacting with a thermal field near a planar medium depicting its: (b) quantized center-of-mass decoherence and (c) increase in momentum uncertainty. }
    \label{Fig: Sch}
\end{figure*}

For the classical center-of-mass motion of a polarizable particle interacting with a thermal EM field, the fluctuations of the field lead to an increase in the center-of-mass momentum uncertainty and a drag force, the two being connected via the fluctuation-dissipation relation~\cite{Einstein_Hopf1910,Mkrtchian2003} (see Fig.~\ref{Fig: Sch}). On the other hand, the quantized center of mass of a particle prepared in a spatial superposition loses its coherence in the presence of thermal fields, as thermal photons scattering off the particle ``measure'' its position~\cite{Joos1985, schlosshauer2007decoherence}.  In free space, the quantized center-of-mass decoherence rate arising from thermal photons can be related to the increase in the momentum uncertainty of its classical center-of-mass, as was pointed out in~\cite{PWM_KS22}.

In this paper, we generalize such a relation by investigating the Brownian motion of the quantum and classical center-of-mass of a neutral, polarizable particle interacting with a thermal field in the presence of a medium. While thermal drag associated with the classical center-of-mass motion of a particle near media has garnered much theoretical interest~\cite{Golyk2013, Oelschlager22, Milton2020}, understanding and mitigating decoherence near media is critical to preserving coherence of matter waves in matter-wave interferometry experiments~\cite{Arndt2022,Hornberger2003, Gerlich2011,Hackermuller2004, Gerlich2011, Mazumdar2020, Fragolino2024}.
Fluctuation-induced decoherence of point charges and charge distributions next to a surface has been studied theoretically \cite{Villanueva2003,Da-Shin2006,Howie2011, Scheel2012, Martinetz2022, Velasco2024} and experimentally \cite{Hasselbach2007, Beierle2018, Kerker2020}. In this work we specifically focus on the decoherence of a neutral, polarizable particle with an \textit{induced dipole}.
  
  We first derive a Quantum Brownian motion (QBM) master equation describing the fluctuation-induced dynamics of the quantized center-of-mass of the particle near a medium~\cite{Sinha2020PRA}.
In the long-wavelength limit, wherein the characteristic coherence length associated with the particle is much smaller than the thermal wavelength~\cite{Joos1985, schlosshauer2007decoherence}, we analyze the resulting thermal Casimir-Polder potential, dissipation and decoherence.
While thermal fluctuations decohere the particle's quantized center of mass in the position basis at a rate $ \Lambda$, the same field fluctuations lead to an increase in the momentum uncertainty for its classical center-of-mass motion, characterized by the momentum diffusion rate $D$. We demonstrate that these two rates are inextricably linked via the relation $ D = 2 \hbar^2 \Lambda$, similar to the free space case~\cite{PWM_KS22}. Furthermore, the quantized center-of-mass dissipation and decoherence are related to the thermal drag and momentum fluctuations via the fluctuation-dissipation relation.


The rest of the paper is organized as follows.  In Section~\ref{Sec:Model}, we describe the model of a neutral, polarizable particle interacting with the quantized EM field in the vicinity of a medium. In Section~\ref{Sec:QBM}, we derive a master equation for the quantized center-of-mass motion that describes the fluctuation-induced Casimir-Polder potential as well as dissipation and decoherence of the center of mass for a polarizable particle near a general medium. In Section~\ref{Sec:Plane}, we analyze the decoherence and dissipation for the specific case of planar motion of the particle near a planar medium.  In Section  \ref{sec:Fokker-Planck} we derive a classical Fokker-Planck--type phase-space equation from the master equation in Section \ref{Sec:QBM} and establish the link between decoherence and dissipation on one side (quantized center of mass) and momentum diffusion and drag on the other side (classical center of mass).  Section~\ref{Sec:CBM} discusses the momentum diffusion and thermal drag experienced by the classical center-of-mass motion of the particle immersed in thermal radiation. We summarize and discuss our findings in Section~\ref{Sec:Disc}.

\section{Model}
\label{Sec:Model}

Let us consider a linearly polarizable particle interacting with the quantized EM field in the presence of a medium, as shown in Fig.~\ref{Fig: Sch}. The total Hamiltonian describing the system is given by $\hat H = \hat H_{M} + \hat H_F + \hat H_I $,  where $ \hat H_M = \frac{\hmb{p}^2}{2M}$ is the Hamiltonian governing the free dynamics for the particle's quantized center-of-mass position $ \hmb{r}$ and momentum $ \hmb{p}$,    $ M$ being the mass of the particle.

\eqn{\hat H_F = \int \dd\omega  \int \dd^3 r' \sum_{\lambda = e,m}\hbar\omega\hmb{f}_\lambda^\dagger \bkt{\mb{r}',\omega}\cdot\hmb{f}_\lambda\bkt{\mb{r}',\omega}
}
represents the Hamiltonian for the quantized EM field in the presence of a medium as described in the macroscopic Quantum Electrodynamics (QED) formalism \cite{Buhmann1, Buhmann2}.  The operators $ \hmb{f}^{(\dagger) }_\lambda\bkt{\mb{r}',\omega}$ represent the creation and annihilation operators for the polaritonic excitations of the field+matter system in the presence of media at frequency $\omega$, position $ \mb{r}'$, with $ \lambda = e,m$ characterizing the quantum noise polarization $(e)$ or magnetization $(m)$. These operators obey the canonical commutation relations $ \sbkt{\hat f_\lambda\bkt{\mb{r}, \omega}_i,\hat f_{\lambda'}^\dagger\bkt{\mb{r}', \omega'}_j }=\delta (\omega - \omega')\delta \bkt{\mb{r} - \mb{r}'}\delta_{\lambda\lambda'}\delta_{ij}$, where the indices $i,j=\cbkt{x,y,z}$ refer to the spatial directions.

The interaction Hamiltonian  between the particle and the EM field is:

\eqn{
\hat H_I = - \int_{V, \mb{r}} \dd^3 s\hat{\mbc{P}}\bkt{\mb{s}}\cdot \hat{\mb{E}}\bkt{\mb{s}},
}
where $\mb{s}$ is a point in the interior of the polarizable particle, centered around position $ \mb{r}$  and integrated over its volume $V$. $\hmb{E} (\mb{s})$ is the electric field and $\hat{\mbc{P}} (\mb{s})$ is the polarization density induced by the field.

The quantized electric field near a surface is given by~\cite{Buhmann1}:
\eqn{\label{Erg}\hat{\mb{E}}\bkt{\mb{r}} =\int\dd\omega\int \dd^3 r'\sum_{ \lambda = e,m}\sbkt {\dbar {G}_\lambda \bkt{\mb{r}, \mb{r}', \omega}\cdot \hat{\mb{f}}_{\lambda}\bkt{\mb{r}',  \omega}\right.&\non\\
\left.+ \hat{\mb{f}}_\lambda^\dagger\bkt{\mb{r}',\omega}\cdot \dbar{G}^\dagger_\lambda \bkt{\mb{r} , \mb{r}', \omega}}&,
}
where the coefficients $ \dbar G_\lambda (\mb{r}, \mb{r}' ,\omega)$ are proportional to the Green tensor  $ \dbar G\bkt{\mb{r}, \mb{r}',\omega}$ of the EM field,  accounting for the presence of the surface~\cite{Buhmann1, Jackson} (see Appendix~\ref{App:GreensTensor} for details).

 We consider the size of the particle to be small compared to the characteristic wavelengths of the EM field, such that it can be described as point-like.  The  total polarization induced in the particle by the above electric field is thus  $ \hat{\mb{P}}\bkt{\mb{r}}\equiv \int_{V, \mb{r}} \dd^3 s \hat{\mbc{P }}\bkt{\mb{s}}\approx V \hat{\mbc{P}}\bkt{\mb{r}}$, such that:
\eqn{\label{Pfg} \hat{\mb{P}}\bkt{\mb{r}} =\int\dd\omega&\int \dd^3 r'\sum_{ \lambda = e,m}\sbkt{  \alpha\bkt{\omega }\dbar {G}_\lambda \bkt{\mb{r}, \mb{r'}, \omega}\cdot \hat{\mb{f}}_{\lambda}\bkt{\mb{r'},  \omega}\right. \non\\
&\left.+ \alpha^\ast \bkt{\omega}  \hat{\mb{f}}_\lambda^\dagger\bkt{\mb{r}',\omega}\cdot \dbar{G}^\dagger_\lambda \bkt{\mb{r} , \mb{r}', \omega} },}
where $ \alpha\bkt{\omega } $ is the linear polarizability of the particle, which we will further assume to be isotropic.

We expand the position of the particle  $\hat{\mb{r}}_p = \mb{r} _0 + \hat{\mb{r} }$ around the classical center-of-mass position $ \mb{r}_0 $ up to first order in quantized center-of-mass displacement $ \hmb{r}$. We obtain $\hat H_I \approx  \hat H_I ^{(1) } +  \hat H_I ^{(2) } $, with:

\eqn{
\hat H_{I}^{(1)}=& - \frac{1}{2}\hat{\mb{P}}\bkt{ \mb{r}_0 }\cdot \hat{\mb{E}}\bkt{ \mb{r}_0}\\ \hat H_{I}^{(2)}=& 
- \frac{1}{2} \hat{\mb{r}}\cdot\mathbf{\nabla}\sbkt{\hat{\mb{P}}\bkt{ \mb{r}_0 }\cdot \hat{\mb{E}}\bkt{ \mb{r}_0}}.
}
Here we assume the long-wavelength limit for interaction between the particle and thermal field, where $\lambda_\mr{th}=4\pi\hbar c/k_B T\gg \Delta r$ with $\lambda_\mr{th}$ being the characteristic thermal photon wavelength at temperature T and $\Delta r$ being the spread of the particle wavefunction in the $x$-$y$ plane ~\cite{schlosshauer2007decoherence}. Thus, the thermal field does not vary appreciably over the range of the wavefunction\footnote{This assumption precludes the short-wavelength limit of the radiation where one would observe a diffusive Brownian motion for the particle's center-of-mass.}. The first term $(\hat H_I^{(1)})$ describes the interaction between the radiation and the induced polarization of the particle, which in turn affects the particle's \textit{classical} center-of-mass. The second term  $(\hat H_I^{(2)})$ describes the interaction between the  \textit{quantized} center-of-mass motion, EM field, and the particle's polarization altogether.

\section{Quantized Center-of-Mass Dynamics}

\subsection{QBM master equation}
\label{Sec:QBM}

We now describe the quantized center-of-mass dynamics resulting from the interaction between the particle and fluctuations of the EM field. The Hamiltonian $\hat H_I^{(2)}$ for the interaction between the particle's center of mass and the fluctuations of the field can be written in the interaction picture as:
\eqn{\tilde H_I^{(2)} \bkt{t}\equiv &e^{ -i \bkt{\hat H_M + \hat H_F}t/\hbar}\hat H_I^{(2)} e^{ i \bkt{\hat H_M + \hat H_F}t/\hbar} \non\\
=& - \tilde {\mb{r}}(t)\cdot\tilde{\mbc{B}}(t),
\label{eq:HI2int}
}
where we define the operators $\tilde {\mb{r}}(t) \equiv e^{-i \hat H_M t/\hbar} \hmb{r} e^{i \hat H_M t/\hbar}$ and  

\eqn{ \label{Eq:Bop}\tilde {\mbc{B}} (t)&\equiv e^{-i \hat H_F t/\hbar} \hat{\mbc{B}} e^{i\hat H_F t/\hbar}=\non\\
&= \frac{1}{2}\sbkt{ \partial_i \hmb{P }\bkt{\mb{r}_0, t}\cdot \hmb{E }\bkt{\mb{r}_0, t} + \hmb{P }\bkt{\mb{r}_0, t}\cdot \partial_i \hmb{E }\bkt{\mb{r}_0, t}}
} 
with $ \hat {\mbc {B}} = \sbkt{\frac{1}{2} \mathbf{\nabla}\bkt{\hat{\mb{P}}\bkt{ \mb{r}_0 }\cdot \hat{\mb{E}}\bkt{ \mb{r}_0}}}$ as the bath operator.
The dynamics of the center-of-mass density matrix $\hat\rho_M=\int d^2r\int d^2r'\rho_M(\mb{r},\mb{r}',t)\ket{\mb{r}}\bra{\mb{r}'} $, with $ \cbkt{\mb{r}, \mb{r'}}$ pertaining to motion of the particle around $\mb r_0$, is  given by the second-order Born-Markov master equation~\cite{BPBook}:
\eqn{
\label{eq:Master_eq}
&\der{\hat\rho_M \bkt{t}}{t} = -\frac{i}{\hbar}\tr_B\sbkt{\tilde {H}_I^{(2)}\bkt{t},\hat\rho_M(0)\otimes \hat\rho_B }\non\\
&-\frac{1}{\hbar^2}\tr_B \int_0 ^\infty \dd\tau \sbkt{\tilde {H}_I^{(2)}\bkt{t},\sbkt{\tilde {H}_I^{(2)}\bkt{t-\tau},\hat\rho_M\bkt{t}\otimes \hat\rho_B }}
}
The first term in Eq. (\ref{eq:Master_eq}) is typically neglected on account of representing a physically irrelevant modification to the system Hamiltonian; here it corresponds to the  Casimir-Polder force along the direction of motion~\cite{CP1948}.

We can simplify the above master equation to obtain the QBM master equation in the Born-Markov approximation. After transforming back to the Schr\"odinger picture we get:
\begin{widetext}

\eqn{\label{Eq:ME1}
\der{\hat\rho_M(t)}{t}= &-\frac{i}{\hbar}\sbkt{\hat{H}_M,\hat\rho_M(t)} -\frac{i}{\hbar}\sum_iC^{(1)}_i\sbkt{\hat r_i,\hat\rho_M(0)}-\frac{1}{2\hbar^2} \int_0 ^\infty\dd \tau \sum_{ij}\sbkt{i \mc{D}_{ij} (\tau ) \sbkt{\hat r_i, \cbkt{\hat r_j, \hat\rho _M(t)}}\right.\non \\ 
&\left. -i \mc{D}_{ij}\bkt{\tau } \frac{\tau }{M}\sbkt{\hat r_i, \cbkt{\hat {p}_j , \hat\rho_M(t)}}- \mc{N}_{ij} (\tau ) \sbkt{\hat r_i, \sbkt{\hat r_j, \hat\rho_M(t) }}+ \mc{N}_{ij}\bkt{\tau } \frac{\tau}{ M} \sbkt{\hat r_i, \sbkt{\hat {p}_j , \hat\rho_M(t)}}},
}

\end{widetext}
where the indices $i,j$ indicate displacements and momenta along the respective axes. The first-order Casimir-Polder force, as well as the dissipation and noise kernels, are defined via the expectation value and two-time correlation functions of the bath operator $ \tilde {\mbc{B}}(t) $ as:
\eqn{\label{C1} C^{(1)}_i\equiv& \avg{{\tilde{\mc{B}}}_i\bkt{t}}\\
\label{diss}\mc{D}_{ij}\bkt{\tau} \equiv& i \avg{\sbkt{{\tilde{\mc{B}}}_i\bkt{t}, {\tilde{\mc{B}}}_j\bkt{t-\tau}}}\\
\label{noise}\mc{N}_{ij}\bkt{\tau }\equiv&\avg{\cbkt{{\tilde{\mc{B}}}_i\bkt{t}, {\tilde{\mc{B}}}_j\bkt{t-\tau}}}.
}
 Ignoring the last term in Eq.~\eqref{Eq:ME1} under the Markov approximation, the master equation simplifies to:
\eqn{
\label{meapprox}
&\der{\hat\rho_M}{t}  \approx -\frac{i}{\hbar}\sbkt{\hat{H}_M,\hat\rho_M(t)}-\frac{i}{\hbar}\sum_iC^{(1)}_i\sbkt{\hat r_i,\hat\rho_M(0)}-\non\\
&\frac{i}{\hbar} \sum_{ij}C_{ij}^{(2)}\sbkt{\hat{r}_i\hat r_j, \hat\rho_M(t)} 
-\frac{i}{\hbar} \sum_{ij}\Gamma_{ij}\sbkt{\hat r_i , \cbkt{\hat {p}_j , \hat\rho_M(t)}}-\non\\
&\sum_{ij}\Lambda_{ij}  \sbkt{\hat r_i, \sbkt{\hat r_j, \hat\rho_M(t) }}.
}
The first term corresponds to the unitary evolution under the free-particle Hamiltonian. The second term corresponds to the first-order  Casimir-Polder potential;  the third term represents a harmonic potential resulting from the second-order Casimir-Polder interaction between the particle and medium~\cite{Jakubec2024}, with  \eqn{\label{Eq:Cij2}C_{ij}^{(2)}\equiv\frac{1}{2\hbar} \int_0 ^\infty \dd\tau\mc{D}_{ij}\bkt{\tau}.}

The fourth term represents dissipation or friction on the quantized center of mass with~\cite{BPBook}:
\eqn{\label{Eq:gammaij}\Gamma_{ij}\equiv \frac{1}{2M\hbar }\int_0 ^\infty \dd\tau \,\tau \mc{D}_{ij}\bkt{\tau} .}
The last term stands for decoherence that acts to localize the center of mass in the position basis, with a localization rate~\cite{BPBook}:
\eqn{\label{Eq:lambdaij}\Lambda_{ij}\equiv \frac{1}{2\hbar^2}\int_0 ^\infty \dd\tau \mc{N}_{ij}\bkt{\tau}.}

We will now investigate these effects for the specific example of a polarizable particle near a planar medium. 

\subsection{Planar medium}
\label{Sec:Plane}
While the QBM master equation (Eq.~\eqref{eq:Master_eq}) describes motion of a polarizable particle interacting with a thermal field near an arbitrary medium, as a concrete example, we consider the motion of the particle in the $x$-$y$ plane ($\hmb{r}=\hat{x}\mb{e}_x+\hat{y}\mb{e}_y$ ) near a planar medium as depicted in Fig.~\ref{Fig: Sch}. In this case, we utilize the statistical independence of the electric field $\hat E \bkt{\mb{r}}$ and its spatial derivative $\partial_i \hat E \bkt{\mb{r}}$ for a thermal state of the EM field~\cite{PWM_KS22}:
\eqn{\label{Eq:EdE}
 \avg{\hat E_p (\mb{r}_0 , t)\partial_i \hat E_k (\mb{r}_0 , t- \tau )}=0,
}
which allows one to simplify each term of the master equation  Eq.~\eqref{meapprox} as follows.

The first term in Eq.~\eqref{meapprox} corresponds to the Casimir-Polder force along direction $i\in \cbkt{x,y}$. Since near a planar surface, the force depends only on the vertical distance $z$ from the surface, the corresponding forces parallel to the surface are zero: $ \avg{\hat E_p (\mb{r})\partial_i \hat E_k (\mb{r})}\rightarrow 0$ \cite{Buhmann1}.

The coefficients $C_{ij}^{(2)}$, $ \Gamma_{ij}$ and $ \Lambda_{ij}$ as defined in Eqs.~\eqref{Eq:Cij2}--\eqref{Eq:lambdaij}, simplify to (see Appendix~\ref{App:CP-potentials} and \ref{App:ME} for details): 
\begin{widetext}
    
\eqn{\label{Eq:CP2}
C_{ij}^{(2)}=&\frac{\hbar \mu_0^2}{2\pi}\sum_{pk=\cbkt{x,y,z}}\int_0^{\infty}d\omega\omega^4\abs{\alpha(\omega)}^2\bkt{2n_\mr{th}(\omega)+1} \non\\
&\sbkt{\re \cbkt{ \partial_i\dbar G_{pk}\bkt{\mb{r}_0 ,\mb{r}_0 , \omega }\partial_j}\im \cbkt{ \dbar G_{pk}\bkt{\mb{r}_0 ,\mb{r}_0 , \omega }}+\re \cbkt{ \dbar G_{pk}\bkt{\mb{r}_0 ,\mb{r}_0 , \omega }}\im \cbkt{ \partial_j\dbar G_{pk}\bkt{\mb{r}_0 ,\mb{r}_0 , \omega }\partial_i}}\\
\label{Eq:gammai}
\Gamma_{ij} =& \frac{\hbar \mu_0 ^2 }{\pi M}\int_0 ^\infty  \dd\omega {\omega}^4 \abs{\alpha\bkt{\omega}}^2\pard{n_\mr{th}\bkt{\omega}}{\omega} \sum_{p,k = \cbkt{x,y,z}} \im \sbkt{ \dbar G_{pk}\bkt{\mb{r}_0 ,\mb{r}_0 , \omega }}\im \sbkt{\partial_i\dbar G_{pk}\bkt{\mb{r}_0 ,\mb{r}_0 , \omega }\partial_j,}\\
\Lambda_{ij}  =& \frac{2\mu_0 ^2 }{\pi } \int \dd\omega \omega^4 \abs{\alpha(\omega)}^2 n_\mr{th} \bkt{\omega}\bkt{n_\mr{th} \bkt{\omega} + 1} \sum_{p,k=\cbkt{x,y,z}}\im\sbkt{ \dbar G _{pk}\bkt{\mb{r}_0 , \mb{r}_0 , \omega}} \im \sbkt{\partial_i\dbar G_{pk}\bkt{\mb{r}_0 , \mb{r}_0 , \omega}\partial_j},
\label{Eq:lambdai}
}
 \end{widetext}
where $n_\mr{th}\bkt{\omega}=\frac{1}{e^{\hbar \omega/k_B T}-1}$ is the Bose-Einstein distribution at temperature $T$, and $ \partial_i \dbar G \bkt{\mb{r}_1 , \mb{r}_2 , \omega}\partial _j$ represents the  double derivative of the Green tensor $ (\dbar G \bkt{\mb{r}_1 , \mb{r}_2 , \omega})$ with respect to $ {r}_{1i} $ and $ r_{2j}$. 

We note a few salient features of the above expressions for the center-of-mass decoherence and dissipation:
 \begin{enumerate}
 \item {For $n_\mr{th}$ given by the Bose-Einstein distribution $ n_\mr{th}\bkt{\omega} = \frac{1}{e^{\hbar \omega /(k_B T) } -1}\implies \pard {n_\mr{th}\bkt{\omega}}{\omega} = \frac{\hbar}{k_B T} n_\mr{th}(\omega)\bkt{n_\mr{th}(\omega) + 1}$, which yields (from Eq.\eqref{Eq:lambdai} and \eqref{Eq:gammai}):
\eqn{
2M\Gamma_{ij} k_B T  = \hbar^2\Lambda_{ij} .
}
in accordance with the fluctuation-dissipation relation between the quantized center-of-mass decoherence and dissipation, as shown in Fig.\ref{Fig: Sch}(a).
Conversely, instead of assuming $n_\mr{th}$ to be the Bose-Einstein distribution, one could use the fluctuation-dissipation relation to relate the dissipation and decoherence coefficients, thus solving for $n_\mr{th}$, and deriving the Bose-Einstein distribution for the field near a medium. While the Planck spectrum can generally assume a different form near boundaries~\cite{Baltes1974}, we show that for the system under consideration, the overall energy spectrum of the field changes due to its field mode distribution, not the thermal state statistics ($n_\mr{th}\bkt{\omega}$)~\cite{PWM_KS22}.
}

\item{Due to planar  symmetry, the  dissipation and decoherence coefficients can be simplified to:
\eqn{
\Gamma_{ij}=&\Gamma\delta_{ij}=\bkt{\Gamma^\mr{free}+\Gamma^\mr{s}}\delta_{ij},\\
\Lambda_{ij}=&\Lambda\delta_{ij}=\bkt{\Lambda^\mr{free}+\Lambda^\mr{s}}\delta_{ij},
} where $\Lambda^\mr{free},\Gamma^\mr{free}$ are the contributions due to free space scattering processes and $\Lambda^\mr{s},\Gamma^\mr{s}$ are the surface-induced contributions, arising from free space and scattering parts of the total Green tensor (see Appendix~\ref{App:GreensTensor}). 
$\Lambda^\mr{free}$, which corresponds to the decoherence due to thermal radiation in the absence of any media, becomes:
\eqn{
&\Lambda^\mr{free} = \non\\
&\frac{1}{18\pi^3 \epsilon_0 ^2 c^8 }\int_0 ^\infty \dd\omega \omega^8 \abs{\alpha(\omega)}^2 n_\mr{th}(\omega)\bkt{n_\mr{th}(\omega)+ 1},
}
in agreement with the previous expressions of free-space decoherence in thermal fields obtained via the scattering theory formalism~\cite{Joos1985, schlosshauer2007decoherence, PWM_KS22}.}

\item{Decoherence $\Lambda_{ij} $ (Eq.~\eqref{Eq:lambdai}) can be connected to the decoherence for an oscillating dipole as derived in ~\cite{Martinetz2022}, by replacing the permanent dipole with an induced one.  }

\item {The center-of-mass decoherence and dissipation vanish at zero temperature, implying that there is no dissipative dynamics purely from quantum fluctuations in the absence of thermal photons, to second-order in the particle-field interaction. For finite temperature, the decoherence and dissipation increase with medium losses determined by $\im \dbar G\bkt{\mb{r}_0, \mb{r}_0, \omega }$.}

\item{Lastly, the expressions of dissipation and decoherence scale with $ \sim n^2_\mr{th} (\omega) + n_\mr{th}\bkt{\omega}$. The $ n_\mr{th}\bkt{\omega}$ and $ n^2_\mr{th}\bkt{\omega}$ terms correspond respectively to the `particle-like' and `wave-like' contributions of the field~\cite{Milonni2019}.
}
 \end{enumerate}

We now examine the link between the  QBM master equation for a polarizable particle near a general medium and its classical center-of-mass motion. 
\subsection{Fokker-Planck Equation}
\label{sec:Fokker-Planck}

To illustrate how decoherence and dissipation ($ \Lambda_{ij}$ and $ \Gamma_{ij} $) are linked to Brownian motion for the classical center of mass, we now derive a phase-space equation from the master equation. To this end, we first write Eq.~\eqref{meapprox} in the position basis as:
\begin{widetext}
\eqn{
\label{eq:meapprox-posrep}
\der{\rho_M(\mb r, \mb r',t)}{t}=&\frac{i\hbar}{2M}\bkt{\frac{\partial^2}{\partial r_i^2}-\frac{\partial^2}{\partial r_i^{'2}}}\rho_M(\mb r,\mb r_i',t)-\frac{i}{\hbar}\sum_{i}C_{i}^{(1)}\bkt{r_i-r_i'}\rho_M(\mb r, \mb r',0)-\frac{i}{\hbar}\sum_{ij}C_{ij}^{(2)}\bkt{r_ir_j-r_i'r_j'}\rho_M(\mb r, \mb r',t)-\non \\
&\sum_{ij}\Gamma_{ij}\bkt{r_i-r_i'}\bkt{\frac{\partial}{\partial r_j}-\frac{\partial}{\partial r_j'}}\rho_M(\mb r, \mb r',t)-\sum_{ij}\Lambda_{ij} \bkt{r_i-r_i'}\bkt{r_j-r_j'}\rho_M(\mb r, \mb r',t)
}
\end{widetext}
Now we take the Wigner-Weyl transform of the above equation by making the substitution $r_i=\bar{r}_i+\frac{a_i}{2}$ and $r_i'=\bar{r}_i-\frac{a_i}{2}$ and using $W(\mb r, \mb p)= \int d\mb a e^{i\mb p \cdot \mb a}/\hbar\rho(\mb r +\mb a/2,\mb r-\mb a/2)$~\cite{BPBook, schlosshauer2007decoherence}. After relabeling $\bar{r}_i$ as $r_i$ we obtain the Quantum Fokker-Planck equation:

\eqn{\label{Eq:FP}
\frac{\partial W}{\partial t}=&-\sum_{i} \frac{p_i}{M} \frac{\partial W}{\partial r_i}-\sum_{i}C_{i}^{(1)}\frac{\partial W}{\partial p_i}-2\sum_{ij}C_{ij}^{(2)}r_i\frac{\partial W}{\partial p_j}\non \\
&  +\sum_{ij}2\Gamma_{ij}\frac{\partial(p_jW)}{\partial p_i}+ \sum_{ij}\Lambda_{ij} \hbar^2 \frac{\partial W}{\partial p_i \partial p_j}. 
}
Analogous to the master equation Eq. (\ref{meapprox}), the second and the third terms correspond to the first- and second-order Casimir-Polder potentials.
The fourth term corresponds to friction and the fifth term represents the momentum diffusion coefficient of the quantized center of mass, which is therefore related to the decoherence coefficient in Eq. (\ref{meapprox}) as:
\eqn{
2 \hbar^ 2 \Lambda_{ij}=\frac{ \avg{\Delta \hat p_i \Delta \hat p_j}_q}{\Delta t} ,
}
where the subscript `$q$' refers to a quantum expectation value. The above relation can be seen as a manifestation of the complementarity of the position and momentum variables. Such a relation was established by ~\cite{Halliwell1996} for general QBM dynamics.

Classically the diffusion coefficient is defined as: 

\eqn{
D_{ij}= \frac{ \avg{\Delta p_i \Delta p_j}_c}{\Delta t}, 
}
where the subscript `$c$' denotes a classical average. Using the Ehrenfest theorem \cite{Ehrenfest1927} to form a correspondence between classical and quantum expectation values, establishes a relation between momentum diffusion for the classical center of mass and decoherence of the quantized center of mass:

\eqn{
\label{eq:relation-pdiff-deco}
\frac{ \avg{\Delta p_i \Delta p_j}_c}{\Delta t}=\frac{ \avg{\Delta \hat p_i \Delta \hat p_j}_q}{\Delta t}\implies D_{ij} =2 \hbar^ 2 \Lambda_{ij}.
}
In essence, this relation confirms that both decoherence (the measurement of the system by the environment) and momentum diffusion arise due to the scattering of thermal photons.
Although Eq.~\eqref{Eq:lambdai} and Eq.~\eqref{Eq:gammai} are derived under the assumption of a planar medium, all the conclusions drawn in this section hold for an arbitrary macroscopic body. 

We can motivate the result $D_{ij} =2 \hbar^ 2 \Lambda_{ij}$ (Eq.~\eqref{eq:relation-pdiff-deco}) through a back-of-the-envelope calculation, treating momentum diffusion as the cumulative effect of many random ``kicks'' delivered to a Brownian particle.
These kicks produce momentum changes $\cdots \delta{\bf p}_{n-1}, \, \delta{\bf p}_{n}, \, \delta{\bf p}_{n+1} \cdots$, which sum to $\Delta{\bf p} \equiv \sum_n \delta{\bf p}_n$ over a time $\Delta t$.
For sufficiently many kicks, $\Delta{\bf p}$ becomes a random sample from the Gaussian distribution
\eqn{
\label{eq:eta}
\eta(\Delta{\bf p}) = \sqrt{\frac{\vert K\vert}{2\pi \Delta t}} \exp\left[ -\frac{ \sum_{ij} K_{ij} \Delta p_i \Delta p_j }{2\Delta t} \right] \, ,
}
where $K\equiv D^{-1}$ is the inverse of the momentum diffusion matrix given by $D_{ij} = \avg{\Delta p_i \Delta p_j}/\Delta t$.

To evaluate the decoherence produced by these kicks, note that when a quantum system undergoes a sudden momentum change $\delta{\bf p}$, its density matrix $\hat\rho$ transforms as
$
\hat\rho \rightarrow e^{-i\delta{\bf p}\cdot \hat{\bf r}/\hbar} \, \hat\rho \, e^{i\delta{\bf p}\cdot \hat{\bf r}/\hbar} \, .
$
Many such changes, summing to $\Delta{\bf p}$ over time $\Delta t$, produce the change (in the position representation)
\eqn{
\rho({\bf r},{\bf r}^\prime,t+\Delta t) = e^{-i\Delta{\bf p}\cdot({\bf r}-{\bf r}^\prime)/\hbar} \rho({\bf r},{\bf r}^\prime,t) \, ,
}
with $\Delta{\bf p} \equiv \sum_n \delta{\bf p}_n$.
Averaging over $\eta(\Delta{\bf p})$ (Eq.~\eqref{eq:eta}) leads to the expression
\eqn{
\label{eq:decoKicks}
\rho({\bf r},&{\bf r}^\prime,t+\Delta t) = \left\langle e^{-i\Delta{\bf p}\cdot({\bf r}-{\bf r}^\prime)/\hbar} \right\rangle \rho({\bf r},{\bf r}^\prime,t) \non\\
&= e^{- \sum_{ij} D_{ij} (r_i-r_i^\prime) (r_j-r_j^\prime)
\Delta t/2\hbar^2} \rho({\bf r},{\bf r}^\prime,t) \, ,
}
describing decoherence in the position representation.

Now consider the effects of the decoherence term in Eq.~\eqref{eq:meapprox-posrep}:
\eqn{
\label{eq:rho-deco}
\frac{\partial}{\partial t} \rho({\bf r},{\bf r}^\prime,t) = - \sum_{ij} \Lambda_{ij} (r_i-r_i^\prime) (r_j-r_j^\prime) \rho({\bf r},{\bf r}^\prime,t) \, .
}
Upon integrating, we obtain the expression
\eqn{
\label{eq:rho-deco}
\rho({\bf r},{\bf r}^\prime,t+\Delta t) = e^{- \sum_{ij} \Lambda_{ij} (r_i-r_i^\prime) (r_j-r_j^\prime) \Delta t} \rho({\bf r},{\bf r}^\prime,t) \, ,
}
which has the same form as Eq.~\eqref{eq:decoKicks}.
Comparing the two gives $D_{ij} =2 \hbar^ 2 \Lambda_{ij}$.

\section{Classical Center-of-Mass Brownian Motion}
\label{Sec:CBM}

We now turn to the Brownian motion properties of the classical center-of-mass of the particle interacting with the thermal field near a surface.
The force acting on the classical center-of-mass of the particle along the $i$-axis is derived from the first-order Hamiltonian $\hat H_I^{(1)}$ as:
\eqn{\hat {F }_i= - {\partial_i \tilde H_I^{(1)}} = \tilde {\mc{B}}_i (t) ,
}
which is the same as the $i^\mr{th}$ component of the bath operator in Eq.~\eqref{Eq:Bop}. The operator nature of the force results from the quantized EM field exerting a force on the particle's classical center of mass. The resulting force is averaged over the thermal density matrix of the field,  leading to a net momentum impulse of zero, such that $\avg{\Delta p_i } =- \int_0 ^{\Delta t} \dd\tau \avg{\tilde {\mc{B}} _i (\tau)}=0$.
The corresponding momentum correlator is thus found from the two-time correlation of the forces as:
\eqn{
\avg{\Delta p_i \Delta p_j } = \int_0 ^{\Delta t} \dd\tau_1 \int_0 ^{\Delta t} \dd\tau_2 \avg{\tilde {\mc{B}}_i\bkt{\tau_1}\tilde {\mc{B}}_i\bkt{\tau_2}}. 
}
We see that the momentum diffusion for the particle's classical center-of-mass diffusion can be found as the rate of increase in the momentum uncertainty with time, $D_{ij}\equiv \frac{\avg{\Delta p_i \Delta p_j}}{\Delta t}$, as (see Appendix~\ref{App:MomFl} for details):
\eqn{\label{Eq:momdif}
&D_{ij}= \frac{4\hbar^2\mu_0 ^2}{ \pi} \int_0 ^\infty \dd\omega \omega^4  n_\mr{th}\bkt{\omega}\bkt{n_\mr{th}\bkt{\omega} + 1}\abs{\alpha \bkt{\omega}}^2\non\\
&\sum_{p,k=\cbkt{x,y,z}}\im\sbkt{ \dbar{ G}_{pk} \bkt{\mb{r}_0 , \mb{r}_0 , \omega} } \im \sbkt{\partial_i\dbar{ G }_{pk}\bkt{\mb{r}_0 , \mb{r}_0 , \omega}\partial_j}.
}
As shown in the previous section, the above expression for momentum diffusion is related to the decoherence rate in Eq. \eqref{Eq:lambdai} as  $D_{ij} = 2\hbar^2 \Lambda_{ij} = 2\hbar^2 \Lambda \delta_{ij} $. 

The above momentum diffusion of a polarizable particle moving through blackbody radiation is accompanied by a drag force of the form $\bkt{F_\mr{drag}}_i=-\xi_{ij}v_j$~\cite{EinsteinHopf}, with $v_j$ as the particle's classical center-of-mass velocity, and $\xi_{ij}$ as the friction coefficient. Using the fluctuation-dissipation theorem $D_{ij}=2\xi_{ij}k_BT$, we obtain the friction coefficient as:

\eqn{\label{Eq:xi}
&\xi_{ij}= \frac{2\hbar^2\mu_0 ^2}{k_BT \pi} \int_0 ^\infty \dd\omega \omega^4  n_\mr{th}\bkt{\omega}\bkt{n_\mr{th}\bkt{\omega} + 1}\abs{\alpha \bkt{\omega}}^2\non\\
&\sum_{p,k=\cbkt{x,y,z}}\im\sbkt{ \dbar G_{pk} \bkt{\mb{r}_0 , \mb{r}_0 , \omega}} \im \sbkt{\partial_i\dbar G_{pk} \bkt{\mb{r}_0 , \mb{r}_0 , \omega}\partial_j}.
}

As before, using planar symmetry, the diffusion and drag coefficients, $D_{ij}=\bkt{D^\mr{free}+D^\mr{s}}\delta_{ij}$ and  $\xi_{ij}=\gamma \delta_{ij}= \bkt{\xi^\mr{free}+\xi^\mr{s}}\delta_{ij}$, respectively, can be separated into free-space ($D^\mr{free}, \xi^\mr{free}$) and surface-induced ($D^\mr{s}, \xi^\mr{s}$) components. In the free-space limit, we recover:
\eqn{
&D^\mr{free}=\frac{\hbar^2}{9\pi^3 \epsilon_0 ^2 c^8 }\int_0 ^\infty \dd\omega \omega^8 \abs{\alpha(\omega)}^2 n_\mr{th}(\omega)\bkt{n_\mr{th}(\omega)+ 1}\\
&\xi^\mr{free}=\non\\
&\frac{\hbar^2}{18k_BT\pi^3 \epsilon_0 ^2 c^8 }\int_0 ^\infty \dd\omega \omega^8 \abs{\alpha(\omega)}^2 n_\mr{th}(\omega)\bkt{n_\mr{th}(\omega)+ 1},
}
in agreement with \cite{PWM_KS22}.

\section{Discussion}
\label{Sec:Disc}

To summarize, we establish a connection between the Brownian motion of the quantum and classical center of mass a neutral, polarizable particle interacting with the fluctuations of a thermal EM field near a medium (Fig.~\ref{Fig: Sch}(a)). We derive a second-order Born-Markov QBM master equation for the particle's quantized center of mass near an arbitrary macroscopic body, which encompasses the thermal Casimir-Polder force, position decoherence and dissipation (Eq.~\eqref{meapprox}). We analyze this master equation for the specific case wherein the motion of the particle is restricted to a plane parallel to a planar medium, demonstrating that the decoherence and dissipation of the particle increase with the medium losses (Eq.~\eqref{Eq:lambdai} and \eqref{Eq:gammai}).  Further deriving the Quantum Fokker-Planck equation corresponding to the QBM master equation (Eq.\eqref{Eq:FP}), we link the quantized center-of-mass decoherence $ \Lambda$ to the classical center of mass momentum diffusion $ D$ via the simple relation $ D = 2\hbar^2 \Lambda$. This relation relies on the position decoherence form of the QBM master equation in the `long-wavelength limit', and is a manifestation of the position-momentum complementarity and Ehrenfest theorem. 
Subsequently, we calculate an explicit expression for the momentum diffusion and classical drag of the polarizable particle in the presence of a planar medium (Eq.~\eqref{Eq:momdif} and \eqref{Eq:xi}).

In the free space limit our decoherence expression agrees with  \cite{schlosshauer2007decoherence,PWM_KS22}, and the classical momentum diffusion and thermal drag near surfaces are in agreement with \cite{Golyk2013, Oelschlager22}. The connection between decoherence and diffusion highlighted in Sec. \ref{sec:Fokker-Planck} has been previously pointed out by \cite{ Halliwell1996} for general QBM dynamics.

Loss of spatial coherence of particles near surfaces is pertinent to matter-wave interferometry. Previous works have investigated the effect of fluctuation-induced van der Waals interactions between particles and gratings which causes dephasing~\cite{Brand2015, Arndt2022}. Dephasing or phase averaging also occurs when the system interacts with an ensemble of EM sources~\cite{Schut2024}. This statistical effect occurs as the phases induced on the system state fluctuate due to a changing environment, leading to an apparent loss of coherence after averaging over many runs of the experiment.


In a recent work, it was illustrated that a polarizable particle in uniformly accelerated motion experiences momentum fluctuations in accord with its observed Unruh-Davies temperature~\cite{PWM_KS2024}. The relation between momentum fluctuations and center-of-mass decoherence suggests that a uniformly accelerating particle prepared in spatial superposition should decohere at a rate determined by the effective Unruh-Davies temperature. 
Such decoherence and Brownian motion in accelerated frames of reference can further correspond to decoherence of particles in the vicinity of black hole event horizons via the equivalence principle, a subject of growing recent interest~\cite{biggs2024, DSW20222, Wei2024, Chen2024}.

\section{Acknowledgements}

We gratefully acknowledge Peter W. Milonni for insightful discussions and feedback on the manuscript. K.S. and Chris J. acknowledge support by the John Templeton Foundation under Award No. 62422. K.S. was also supported by the National Science Foundation under Grant No. PHY-2418249.  Clemens J. acknowledges support by the John Templeton Foundation under Award No. 63033. 

\begin{widetext}
\appendix
\section{The Electromagnetic Green tensor}
\label{App:GreensTensor}


The electromagnetic Green tensor $\dbar {G} \bkt{\mb{r}_1 ,\mb{r}_2, \omega} $ represents the propagator for the EM field  in the presence of media, accounting for how a photon at frequency $ \omega$ propagates between positions $ \mb r_1 $ and $ \mb r_2$. The Green tensor can be found as  the solution of the Helmholtz equation~\cite{Jackson},
\eqn{
\bkt{\nabla_1 \times \frac{1}{\mu(\mb{r}_1, \omega) }\nabla_1 \times - \frac{\omega^2 }{c^2}\epsilon\bkt{\mb {r}_1, \omega}}\dbar G (\mb{r}_1, \mb{r}_2 ,\omega)=\mb{\delta }\bkt{\mb{r}_1 - \mb{r_2}},
}
where $\epsilon (\mb {r}, \omega)$ and $\mu(\mb{r}, \omega)$ represent the spatially dependent dielectric permittivity and magnetic permeability in the presence of a  medium. The Green tensor encompasses the effects of surface geometry, optical and material properties of the medium, thereby modifying the density of states of the EM environment. For a general medium the Green tensor can always be split up into a free space Green tensor $\dbar{G}_{\mr{0}} \bkt{\mb{r}_1,\mb{r}_2, \omega}$ and a scattering Green tensor $\dbar{G}_{\mr{s}} \bkt{\mb{r}_1,\mb{r}_2, \omega}$.

The free space Green tensor describes the EM field when $\epsilon(\mb {r}, \omega)=\epsilon_0$ and $ \mu(\mb {r}, \omega)=\mu_0$, i.e. when there are no media present, and between the points $\mb{r}_1$ and $\mb{r}_2$ is given as: 

\eqn{
\label{greenfree}
\dbar{G}_{\mr{0}} \bkt{\mb{r}_1,\mb{r}_2, \omega} =
\frac{e^{ikr}}{4\pi k^2 r^3}\cbkt{f(kr)\mathbb{1}-h(kr)\mb{e}_r\otimes\mb{e}_r}
}

where $f\bkt{x}\equiv1-ix-x^2$, 
$h\bkt{x}\equiv3-3ix-x^2$, $r = \abs{\mb{r}_1- \mb{r}_2}$, and $\mb{e}_r=\mb{r}/r$. $\dbar{G}_{\mr{0}} \bkt{\mb{r}_1,\mb{r}_2, \omega}$ includes all the processes that do not involve scattering of any surfaces and is thus independent of any surface properties and only depends in the distance between $\mb{r}_1$ and $\mb{r}_2$.

The surface Green tensor, which describes the response of the electric field in the presence of a planar surface, is given by: 
\eqn{
\dbar{G}_{\mr{s}} \bkt{\mb{r}_1,\mb{r}_2, \omega} =\frac{1}{8\pi}\int_0^\infty dk^\parallel \frac{k^\parallel}{\kappa^\perp}e^{-\kappa^\perp z_+}\sbkt{
\begin{pmatrix}
    J_0(k^\parallel x)+J_2(k^\parallel x) & 0 & 0 \\
    0 & J_0(k^\parallel x)-J_2(k^\parallel x) & 0 \\
    0&0&0
\end{pmatrix}r_s
\right. \non\\ \left. +\frac{c^2}{\omega^2}
\begin{pmatrix}
    \kappa{^\perp2} \sbkt{J_0(k^\parallel x)-J_2(k^\parallel x)} & 0 & 2k^\parallel \kappa^\perp J_1(k^\parallel x)\\
    0 & \kappa^{\perp2} \sbkt{J_0(k^\parallel x)+J_2(k^\parallel x)} & 0 \\
    -2k^\parallel \kappa^\perp J_1(k^\parallel x) & 0 & 2k^{\parallel2} J_0(k^\parallel x)
\end{pmatrix}r_p
}}
where $k^\parallel$, $\kappa^\perp$ are the wavevectors parallel and perpendicular to the surface with $\omega^2/c^2=k^{\parallel2}-\kappa^{\perp2}$, $z_+=z_1+z_2$, $x=x_1-x_2$, $J_0(k^\parallel x)$ and $J_2(k^\parallel x)$ are the cylindrical Bessel function of the zeroth and second kind and $r_s$, $r_p$ are the Fresnel coefficients for s and p-polarized light. 


\section{Casimir-Polder Potentials}
\label{App:CP-potentials}

The interaction between the fluctuating field and the polarization induced by it in the particle, leads to the Casimir-Polder potential. The second-order master equation approach used in this paper yields Casimir-Polder potential up to second order in the polarizability $\alpha(\omega)$. The order refers to the number of scattering events taken into account. First order thus means that no scattered fields are considered, whereas second order means that fields that scatter off the particle once are taken into account. The first-order Casimir-Polder potential is given by~\cite{Buhmann1}: 

\eqn{
\mc{U}_\mr{CP}^{(1)}=-\frac{\hbar\mu_0}{2\pi}\int_0^\infty d\omega \alpha(\omega)\omega^2(2n_\mr{th}(\omega)+1)\tr \sbkt{\im \ \dbar G(\mb{r}_0,\mb{r}_0,\omega)}.
}
The first term in Eq. (\ref{eq:Master_eq}) gives rise to the first-order Casimir-Polder force,  defined as the derivative of the above Casimir-Polder potential along the $i$ axis, i.e., $ C^{(1)}_i \equiv \partial_i \mc{U}_\mr{CP}^{(1)}$.
When the motion of the particle is restricted to the x-y plane, the first derivative of the Casimir-Polder potential is zero because $\mc{U}_\mr{CP}^{(1)}$ only varies in the z-direction via the scattering Green tensor (see App. \ref{App:GreensTensor}). 

The second order Casimir-Polder potential $\mc{U}_\mr{CP}^{(2)}$  is given by \cite{Jakubec2024}: 

\eqn{
\mc{U}_\mr{CP}^{(2)}=\frac{\hbar \mu_0^2}{2\pi}\sum_{pk=\cbkt{x,y,z}}\int_0^{\infty}d\omega\omega^4\abs{\alpha(\omega)}^2\bkt{2n_\mr{th}(\omega)+1} 
\sbkt{\re \ \dbar G_{pk}\bkt{\mb{r}_0 ,\mb{r}_0 , \omega }\im  \ \dbar G_{pk}\bkt{\mb{r}_0 ,\mb{r}_0 , \omega }.
}}
It is a second-order potential and therefore represents processes that involve scattering off the particle. The second term in Eq.~\eqref{meapprox} corresponds to the second derivative of the above potential along $ i$ and $j$ axes, i.e., $ C^{(2)}_{ij} \equiv \partial_i \mc{U}_\mr{CP}^{(2)} \partial_j $.

\section{Noise and dissipation kernels}
\label{App:ME}
\subsection{Noise kernel}

In the following the indices $i,j$ will take the values $i,j=\cbkt{x,y}$ and the indices $p,k$ will take the values $p,k=\cbkt{x,y,z}$.    
    \eqn{\label{eq:Ntau}
\mc{N}_{ij}\bkt{\tau }
=&  \frac{1}{4}\avg{  \sbkt{\partial_i \hmb{P }\bkt{\mb{r}_0, t}\cdot \hmb{E }\bkt{\mb{r}_0, t} + \hmb{P }\bkt{\mb{r}_0, t}\cdot \partial_i \hmb{E }\bkt{\mb{r}_0, t}}\sbkt{\partial_j \hmb{P }\bkt{\mb{r}_0, t- \tau}\cdot \hmb{E }\bkt{\mb{r}_0, t- \tau} + \hmb{P }\bkt{\mb{r}_0, t- \tau}\cdot \partial_j \hmb{E }\bkt{\mb{r}_0, t- \tau}}\right.\non\\
& \left. + \sbkt{\partial_j \hmb{P }\bkt{\mb{r}_0, t- \tau}\cdot \hmb{E }\bkt{\mb{r}_0, t- \tau} + \hmb{P }\bkt{\mb{r}_0, t- \tau}\cdot \partial_j \hmb{E }\bkt{\mb{r}_0, t- \tau}}\sbkt{\partial_i \hmb{P }\bkt{\mb{r}_0, t}\cdot \hmb{E }\bkt{\mb{r}_0, t} + \hmb{P }\bkt{\mb{r}_0, t}\cdot \partial_i \hmb{E }\bkt{\mb{r}_0, t}}}}
\eqn{
=&   \frac{1}{4}\sum_{p,k}\sbkt{\underbrace{\avg{\partial_i \hat{P }_p\bkt{\mb{r}_0, t} \partial_j \hat{P }_k\bkt{\mb{r}_0, t- \tau}}}_\mr{(IA)}\underbrace{ \avg{ \hat{E }_p\bkt{\mb{r}_0, t}  \hat{E }_k\bkt{\mb{r}_0, t- \tau}}}_\mr{(IB)}\right.\non\\
&\left.+\underbrace{ \avg{ \hat{P }_p\bkt{\mb{r}_0, t} \hat{P }_k\bkt{\mb{r}_0, t- \tau}}}_\mr{(IIA)}\underbrace{ \avg{ \partial_i\hat{E }_p\bkt{\mb{r}_0, t} \partial_j \hat{E }_k\bkt{\mb{r}_0, t- \tau}}}_\mr{(IIB)}\right.\non\\
&\left.+\underbrace{ \avg{ \hat{P }_p\bkt{\mb{r}_0, t} \hat{E }_k\bkt{\mb{r}_0, t- \tau}}}_\mr{(IIIA)}\underbrace{ \avg{ \partial_i \hat{E }_p\bkt{\mb{r}_0, t} \partial_j \hat{P }_k\bkt{\mb{r}_0, t- \tau}}}_\mr{(IIIB)}\right.\non\\
&\left.+ \underbrace{\avg{ \partial_i \hat{P }_p\bkt{\mb{r}_0, t} \partial_j \hat{E }_k\bkt{\mb{r}_0, t- \tau}}}_\mr{(IVA)}\underbrace{\avg{ \hat{E }_p\bkt{\mb{r}_0, t} \hat{P }_k\bkt{\mb{r}_0, t- \tau}}}_\mr{(IVB)} \right.\non\\
&\left.+\underbrace{ \avg{\partial_j \hat{P }_p\bkt{\mb{r}_0, t- \tau} \partial_i \hat{P }_k\bkt{\mb{r}_0, t}} }_\mr{(VA)}\underbrace{\avg{ \hat{E }_p\bkt{\mb{r}_0, t- \tau}  \hat{E }_k\bkt{\mb{r}_0, t}}}_\mr{(VB)}\right.\non\\
&\left.+ \underbrace{\avg{ \hat{P }_p\bkt{\mb{r}_0, t- \tau} \hat{P }_k\bkt{\mb{r}_0, t}}}_\mr{(VIA)}\underbrace{ \avg{ \partial_j\hat{E }_p\bkt{\mb{r}_0, t- \tau} \partial_i \hat{E }_k\bkt{\mb{r}_0, t}}}_\mr{(VIB)}\right.\non\\
&\left.+ \underbrace{\avg{ \hat{P }_p\bkt{\mb{r}_0, t- \tau} \hat{E }_k\bkt{\mb{r}_0, t}}}_\mr{(VIIA)}\underbrace{ \avg{ \partial_j \hat{E }_p\bkt{\mb{r}_0, t- \tau} \partial_i \hat{P }_k\bkt{\mb{r}_0, t}}}_\mr{(VIIB)}\right.\non\\
&\left.+ \underbrace{\avg{ \partial_j \hat{P }_p\bkt{\mb{r}_0, t- \tau} \partial_i \hat{E }_k\bkt{\mb{r}_0, t}}}_\mr{(VIIIA)}\underbrace{\avg{ \hat{E }_p\bkt{\mb{r}_0, t- \tau} \hat{P }_k\bkt{\mb{r}_0, t}}  }_\mr{(VIIIB)} } ,
}
where we have used the statistical independence of the electric field $\hat E_p \bkt{\mb{r}_0 ,t}$ and its spatial derivative $\partial_i \hat E_p \bkt{\mb{r}_0 ,t}$ as in Eq.~\eqref{Eq:EdE}. The same applies to the particle polarization $\hat P_p \bkt{\mb{r}_0 ,t}$ $ \bkt{\avg{\hat P_p (\mb{r}_0 , t)\partial_i \hat P_k (\mb{r}_0 , t- \tau )}=0}$.

We simplify each of the above terms as follows:
\eqn{
\mr{(IA)} =& \avg{\partial_i \hat{P }_p\bkt{\mb{r}_0, t} \partial_j \hat{P }_k\bkt{\mb{r}_0, t- \tau}}\\
= & \int \dd\omega_1  \int \dd\omega_2 \int \dd r_1^3\int \dd r_2^3\sum_{\lambda_1 ,\lambda_2} \avg{\sbkt{ \alpha \bkt{\omega_1} \partial_i (\dbar G_1\bkt{\mb{r}_0 }\cdot\hmb{f}_{1})_p e^{-i \omega_1 t} + \alpha ^\ast(\omega_1) \partial_j (\hmb{f}_{1}^\dagger\cdot \dbar G^\dagger_{1}\bkt{\mb{r}_0})_p e^{i \omega_1 t} }\right.\non\\
& \left. \sbkt{ \alpha \bkt{\omega_2} \partial_i (\dbar G_2\bkt{\mb{r}_0 }\cdot\hmb{f}_{2})_k e^{-i \omega_2 (t-\tau)} + \alpha ^\ast(\omega_2) \partial_j (\hmb{f}_{2}^\dagger\cdot \dbar G^\dagger_{2}\bkt{\mb{r}_0})_k e^{i \omega_2 (t-\tau)} }} ,
}
where we have used the shorthand notation $ \dbar {G}_{1,2}^{(\dagger)} (\mb{\omega_0 }) \equiv  \dbar {G}^{(\dagger)}_{\lambda_{1,2}}\bkt{\mb{r}_0, \mb{r}_{1,2}, \omega_{1,2}} $ and $ \hmb{f}^{(\dagger)}_{1,2}\equiv \hmb{f}^{(\dagger)}_{\lambda_{1,2}}\bkt{\mb{r}_{1,2}, \omega_{1,2}}$. To take the average over the thermal density matrix of the EM field bath, we note that $ \avg{\hmb{f}_1\hmb{f}_2^\dagger} =( n_\mr{th}(\omega_1)+ 1) \delta (\omega_1 - \omega_2) \delta (\mb{r}_1 - \mb{r}_2) \delta _{\lambda_1\lambda_2}$,  $ \avg{\hmb{f}_1^\dagger\hmb{f}_2} =n_\mr{th}(\omega_1) \delta (\omega_1 - \omega_2) \delta (\mb{r}_1 - \mb{r}_2) \delta _{\lambda_1\lambda_2}$ and $ \avg{\hmb{f}_1\hmb{f}_2} = \avg{\hmb{f}_1^\dagger\hmb{f}_2^\dagger}=0$. We  thus obtain:

\eqn{
\label{eq:1a}\mr{(IA)} = \frac{\hbar \mu_0 }{\pi }\int \dd\omega \omega^2\abs{\alpha\bkt{\omega}}^2 \im \sbkt{\partial_i \bkt{\dbar G \bkt{\mb{r}_0 , \mb{r}_0 , \omega}}_{pk}\partial_j  }\sbkt{\bkt{n_\mr{th}\bkt{\omega} + 1} e^{-i \omega \tau }+ n_\mr{th}\bkt{\omega}  e^{i \omega \tau }},
}
where we have used the fluctuation-dissipation relation for the Green tensor. Similarly one can simplify the remaining terms:

\eqn{
\label{eq:1b}
\mr{(IB)} = &\frac{\hbar \mu_0 }{\pi }\int \dd\omega \omega^2 \im \sbkt{\bkt{\dbar G \bkt{\mb{r}_0 , \mb{r}_0 , \omega}}_{pk} }\sbkt{\bkt{n_\mr{th}\bkt{\omega} + 1} e^{-i \omega \tau }+ n_\mr{th}\bkt{\omega}  e^{i \omega \tau }}\\
\label{eq:2a}
\mr{(IIA)}=&\frac{\hbar \mu_0 }{\pi }\int \dd\omega \omega^2\abs{\alpha\bkt{\omega}}^2 \im \sbkt{\bkt{\dbar G \bkt{\mb{r}_0 , \mb{r}_0 , \omega}}_{pk}}\sbkt{\bkt{n_\mr{th}\bkt{\omega} + 1} e^{-i \omega \tau }+ n_\mr{th}\bkt{\omega}  e^{i \omega \tau }}\\
\label{eq:2b}
\mr{(IIB)} = & \frac{\hbar \mu_0 }{\pi }\int \dd\omega \omega^2\im \sbkt{\partial_i \bkt{\dbar G \bkt{\mb{r}_0 , \mb{r}_0 , \omega}}_{pk}\partial_j  }\sbkt{\bkt{n_\mr{th}\bkt{\omega} + 1} e^{-i \omega \tau }+ n_\mr{th}\bkt{\omega}  e^{i \omega \tau }}\\
\label{eq:3a}
\mr{(IIIA)} = & \frac{\hbar \mu_0 }{\pi }\int \dd\omega \omega^2\im \sbkt{ \bkt{\dbar G \bkt{\mb{r}_0 , \mb{r}_0 , \omega}}_{pk} }\sbkt{\alpha(\omega)\bkt{n_\mr{th}\bkt{\omega} + 1} e^{-i \omega \tau }+ \alpha^\ast(\omega)n_\mr{th}\bkt{\omega}  e^{i \omega \tau }}\\
\label{eq:3b}
\mr{(IIIB)} = & \frac{\hbar \mu_0 }{\pi }\int \dd\omega \omega^2\im \sbkt{ \partial_i \bkt{\dbar G \bkt{\mb{r}_0 , \mb{r}_0 , \omega}}_{pk} \partial_j }\sbkt{\alpha^\ast(\omega)\bkt{n_\mr{th}\bkt{\omega} + 1} e^{-i \omega \tau }+ \alpha(\omega)n_\mr{th}\bkt{\omega}  e^{i \omega \tau }}\\
\label{eq:4a}
\mr{(IVA)} = & \frac{\hbar \mu_0 }{\pi }\int \dd\omega \omega^2\im \sbkt{\partial_i \bkt{\dbar G \bkt{\mb{r}_0 , \mb{r}_0 , \omega}}_{pk} \partial_j}\sbkt{\alpha(\omega)\bkt{n_\mr{th}\bkt{\omega} + 1} e^{-i \omega \tau }+ \alpha^\ast(\omega)n_\mr{th}\bkt{\omega}  e^{i \omega \tau }}\\
\label{eq:4b}
\mr{(IVB)} = & \frac{\hbar \mu_0 }{\pi }\int \dd\omega \omega^2\im \sbkt{ \bkt{\dbar G \bkt{\mb{r}_0 , \mb{r}_0 , \omega}}_{pk}}\sbkt{\alpha^\ast(\omega)\bkt{n_\mr{th}\bkt{\omega} + 1} e^{-i \omega \tau }+ \alpha(\omega)n_\mr{th}\bkt{\omega}  e^{i \omega \tau }}\\ 
\label{eq:5a}
\mr{(VA)} = &  \frac{\hbar \mu_0 }{\pi }\int \dd\omega \omega^2\abs{\alpha\bkt{\omega}}^2 \im \sbkt{\partial_j \bkt{\dbar G \bkt{\mb{r}_0 , \mb{r}_0 , \omega}}_{pk}\partial_i  }\sbkt{\bkt{n_\mr{th}\bkt{\omega} + 1} e^{i \omega \tau }+ n_\mr{th}\bkt{\omega}  e^{- i \omega \tau }}\\
\label{eq:5b}
\mr{(VB)} = &\frac{\hbar \mu_0 }{\pi }\int \dd\omega \omega^2 \im \sbkt{\bkt{\dbar G \bkt{\mb{r}_0 , \mb{r}_0 , \omega}}_{pk} }\sbkt{\bkt{n_\mr{th}\bkt{\omega} + 1} e^{i \omega \tau }+ n_\mr{th}\bkt{\omega}  e^{-i \omega \tau }}\\
\label{eq:6a}
\mr{(VIA)} =&\frac{\hbar \mu_0 }{\pi }\int \dd\omega \omega^2\abs{\alpha\bkt{\omega}}^2 \im \sbkt{\bkt{\dbar G \bkt{\mb{r}_0 , \mb{r}_0 , \omega}}_{pk}}\sbkt{\bkt{n_\mr{th}\bkt{\omega} + 1} e^{i \omega \tau }+ n_\mr{th}\bkt{\omega}  e^{-i \omega \tau }}\\
\label{eq:6b}
\mr{(VIB)} = & \frac{\hbar \mu_0 }{\pi }\int \dd\omega \omega^2\im \sbkt{\partial_j \bkt{\dbar G \bkt{\mb{r}_0 , \mb{r}_0 , \omega}}_{pk}\partial_i  }\sbkt{\bkt{n_\mr{th}\bkt{\omega} + 1} e^{i \omega \tau }+ n_\mr{th}\bkt{\omega}  e^{-i \omega \tau }}\\
\label{eq:7a}
\mr{(VIIA)} = & \frac{\hbar \mu_0 }{\pi }\int \dd\omega \omega^2\im \sbkt{ \bkt{\dbar G \bkt{\mb{r}_0 , \mb{r}_0 , \omega}}_{pk} }\sbkt{\alpha(\omega)\bkt{n_\mr{th}\bkt{\omega} + 1} e^{i \omega \tau }+ \alpha^\ast(\omega)n_\mr{th}\bkt{\omega}  e^{-i \omega \tau }}\\
\label{eq:7b}
\mr{(VIIB)} = & \frac{\hbar \mu_0 }{\pi }\int \dd\omega \omega^2\im \sbkt{ \partial_j \bkt{\dbar G \bkt{\mb{r}_0 , \mb{r}_0 , \omega}}_{pk} \partial_i }\sbkt{\alpha^\ast(\omega)\bkt{n_\mr{th}\bkt{\omega} + 1} e^{i \omega \tau }+ \alpha(\omega)n_\mr{th}\bkt{\omega}  e^{-i \omega \tau }}\\
\label{eq:8a}
\mr{(VIIIA)} = & \frac{\hbar \mu_0 }{\pi }\int \dd\omega \omega^2\im \sbkt{\partial_j \bkt{\dbar G \bkt{\mb{r}_0 , \mb{r}_0 , \omega}}_{pk} \partial_i}\sbkt{\alpha(\omega)\bkt{n_\mr{th}\bkt{\omega} + 1} e^{i \omega \tau }+ \alpha^\ast(\omega)n_\mr{th}\bkt{\omega}  e^{-i \omega \tau }}\\
\label{eq:8b}
\mr{(VIIIB)} = & \frac{\hbar \mu_0 }{\pi }\int \dd\omega \omega^2\im \sbkt{ \bkt{\dbar G \bkt{\mb{r}_0 , \mb{r}_0 , \omega}}_{pk}}\sbkt{\alpha^\ast(\omega)\bkt{n_\mr{th}\bkt{\omega} + 1} e^{i \omega \tau }+ \alpha(\omega)n_\mr{th}\bkt{\omega}  e^{-i \omega \tau }}
}

Putting all the terms together, we get the noise kernel:

\eqn{
&\mc{N}_{ij} \bkt{\tau} =\frac{\hbar^2 \mu_0 ^2 }{2\pi ^2} \sum_{p,k}\int_0 ^\infty  \dd\omega \omega^2  \int _0 ^\infty \dd\omega' {\omega'} ^2\non\\
& \sbkt{ \im \cbkt{\partial_i\bkt{\dbar G\bkt{\mb{r}_0 ,\mb{r}_0 , \omega }}_{pk}\partial_j}\im \cbkt{ \bkt{\dbar G\bkt{\mb{r}_0 ,\mb{r}_0 , \omega' }}_{pk}} +\im \cbkt{\bkt{\dbar G\bkt{\mb{r}_0 ,\mb{r}_0 , \omega }}_{pk}}\im \cbkt{\partial_j \bkt{\dbar G\bkt{\mb{r}_0 ,\mb{r}_0 , \omega' }}_{pk}\partial_i } }\non\\
& \sbkt{\abs{\alpha\bkt{\omega}}^2\cbkt{\bkt{\bkt{n_\mr{th}\bkt{\omega}  + 1}\bkt{n_\mr{th}\bkt{\omega'}  + 1} + n_\mr{th}\bkt{\omega}n_\mr{th}\bkt{\omega'}  } \cos \bkt{(\omega + \omega') \tau}   \right.\right.\non\\
& \left. \left.+ \bkt{\bkt{n_\mr{th}\bkt{\omega}  + 1}n_\mr{th}\bkt{\omega'}   + n_\mr{th}\bkt{\omega}\bkt{n_\mr{th}\bkt{\omega'}  +1}} \cos \bkt{(\omega - \omega') \tau}  }\right.\non\\
&\left.+  \cbkt{\bkt{\alpha\bkt{\omega}\alpha^\ast\bkt{\omega'}\bkt{n_\mr{th} \bkt{\omega} + 1}\bkt{n_\mr{th} \bkt{\omega'} + 1}+ \alpha^\ast\bkt{\omega}\alpha\bkt{\omega'}n_\mr{th} \bkt{\omega}n_\mr{th} \bkt{\omega'}}}\cos\bkt{(\omega + \omega' )\tau} \right.\non\\
&\left.+ \cbkt{\bkt{\alpha (\omega)\alpha\bkt{\omega'}\bkt{n_\mr{th} (\omega) + 1} n_\mr{th}\bkt{\omega'} + \alpha ^\ast(\omega)\alpha^\ast\bkt{\omega'}n_\mr{th} (\omega) \bkt{n_\mr{th}\bkt{\omega'}  +1}} \cos \bkt{(\omega - \omega')\tau}}}.
}

The decoherence rate as defined in Eq.~\eqref{Eq:lambdaij} is obtained by integrating the  noise kernel over $ \tau$, using $ \int _0 ^\infty \dd\tau \cos \bkt{(\omega - \omega')\tau } = \pi \delta \bkt{\omega - \omega'}$, giving: 

\eqn{
\Lambda_{ij}  = \frac{2\mu_0 ^2 }{\pi } \int \dd\omega \omega^4 \abs{\alpha(\omega)}^2 n_\mr{th} \bkt{\omega}\bkt{n_\mr{th} \bkt{\omega} + 1} \sum_{p,k=\cbkt{x,y,z}}\im\sbkt{ \bkt{\dbar G \bkt{\mb{r}_0 , \mb{r}_0 , \omega}} _{pk}} \im \sbkt{\partial_i\bkt{\dbar G \bkt{\mb{r}_0 , \mb{r}_0 , \omega}}_{pk}\partial_j}
}

Following App. \ref{App:GreensTensor}, we can split up the Green tensor into a free space and a scattering Green tensor. Doing so leaves us with the free space and scattering decoherence coefficients: 

\eqn{
\Lambda^\mr{free}_{ij}  = &\frac{2\mu_0 ^2 }{\pi } \int \dd\omega \omega^4 \abs{\alpha(\omega)}^2 n_\mr{th} \bkt{\omega}\bkt{n_\mr{th} \bkt{\omega} + 1} \sum_{p,k=\cbkt{x,y,z}}\im\sbkt{ \bkt{\dbar G_\mr{0} \bkt{\mb{r}_0 , \mb{r}_0 , \omega}} _{pk}} \im \sbkt{\partial_i\bkt{\dbar G_\mr{0} \bkt{\mb{r}_0 , \mb{r}_0 , \omega}}_{pk}\partial_j}}

\eqn{
\Lambda^\mr{s}_{ij}  = &\frac{2\mu_0 ^2 }{\pi } \int \dd\omega \omega^4 \abs{\alpha(\omega)}^2 n_\mr{th} \bkt{\omega}\bkt{n_\mr{th} \bkt{\omega} + 1} \sum_{p,k=\cbkt{x,y,z}}\im\sbkt{ \bkt{\dbar G_\mr{0} \bkt{\mb{r}_0 , \mb{r}_0 , \omega}} _{pk}} \im \sbkt{\partial_i\bkt{\dbar G_\mr{s}  \bkt{\mb{r}_0 , \mb{r}_0 , \omega}}_{pk}\partial_j}+ \non\\
+&\frac{2\mu_0 ^2 }{\pi } \int \dd\omega \omega^4 \abs{\alpha(\omega)}^2 n_\mr{th} \bkt{\omega}\bkt{n_\mr{th} \bkt{\omega} + 1} \sum_{p,k=\cbkt{x,y,z}}\im\sbkt{ \bkt{\dbar G_\mr{s}  \bkt{\mb{r}_0 , \mb{r}_0 , \omega}} _{pk}} \im \sbkt{\partial_i\bkt{\dbar G_\mr{0} \bkt{\mb{r}_0 , \mb{r}_0 , \omega}}_{pk}\partial_j}+ \non\\
+&\frac{2\mu_0 ^2 }{\pi } \int \dd\omega \omega^4 \abs{\alpha(\omega)}^2 n_\mr{th} \bkt{\omega}\bkt{n_\mr{th} \bkt{\omega} + 1} \sum_{p,k=\cbkt{x,y,z}}\im\sbkt{ \bkt{\dbar G_\mr{s} \bkt{\mb{r}_0 , \mb{r}_0 , \omega}} _{pk}} \im \sbkt{\partial_i\bkt{\dbar G_\mr{s}  \bkt{\mb{r}_0 , \mb{r}_0 , \omega}}_{pk}\partial_j}
}.

$\Lambda^\mr{free}_{ij}$ only depends on the free space Green tensor and represents decoherence due to process that do not involve scattering off the surface. The first and second terms of $\Lambda^\mr{s}_{ij}$ correspond to the field scattered off the surface interacting with the polarization induced by the free space field and vice-versa. The last term of $\Lambda^\mr{s}_{ij}$ describes processes where the field scattered off the surface interacts with the polarization induced by the scattered field.

\subsection{Dissipation kernel}
In the following the indices $i,j$ will take the values $i,j=\cbkt{x,y}$ and the indices $p,k$ will take the values $p,k=\cbkt{x,y,z}$. We can similarly simplify the dissipation kernel as follows:
\eqn{
\mc{D}_{ij } \bkt{\tau } = &  \frac{1}{4}\avg{  \sbkt{\partial_i \hmb{P }\bkt{\mb{r}_0, t}\cdot \hmb{E }\bkt{\mb{r}_0, t} + \hmb{P }\bkt{\mb{r}_0, t}\cdot \partial_i \hmb{E }\bkt{\mb{r}_0, t}}\sbkt{\partial_j \hmb{P }\bkt{\mb{r}_0, t- \tau}\cdot \hmb{E }\bkt{\mb{r}_0, t- \tau} + \hmb{P }\bkt{\mb{r}_0, t- \tau}\cdot \partial_j \hmb{E }\bkt{\mb{r}_0, t- \tau}}\right.\non\\
& \left. - \sbkt{\partial_j \hmb{P }\bkt{\mb{r}_0, t- \tau}\cdot \hmb{E }\bkt{\mb{r}_0, t- \tau} + \hmb{P }\bkt{\mb{r}_0, t- \tau}\cdot \partial_j \hmb{E }\bkt{\mb{r}_0, t- \tau}}\sbkt{\partial_i \hmb{P }\bkt{\mb{r}_0, t}\cdot \hmb{E }\bkt{\mb{r}_0, t} + \hmb{P }\bkt{\mb{r}_0, t}\cdot \partial_i \hmb{E }\bkt{\mb{r}_0, t}}}}

\eqn{=&   
\frac{1}{4}\sum_{p,k}\sbkt{\underbrace{\avg{\partial_i \hat{P }_p\bkt{\mb{r}_0, t} \partial_j \hat{P }_k\bkt{\mb{r}_0, t- \tau}}}_\mr{(IA)}\underbrace{ \avg{ \hat{E }_p\bkt{\mb{r}_0, t}  \hat{E }_k\bkt{\mb{r}_0, t- \tau}}}_\mr{(IB)}\right.\non\\
&\left.+\underbrace{ \avg{ \hat{P }_p\bkt{\mb{r}_0, t} \hat{P }_k\bkt{\mb{r}_0, t- \tau}}}_\mr{(IIA)}\underbrace{ \avg{ \partial_i\hat{E }_p\bkt{\mb{r}_0, t} \partial_j \hat{E }_k\bkt{\mb{r}_0, t- \tau}}}_\mr{(IIB)}\right.\non\\
&\left.+\underbrace{ \avg{ \hat{P }_p\bkt{\mb{r}_0, t} \hat{E }_k\bkt{\mb{r}_0, t- \tau}}}_\mr{(IIIA)}\underbrace{ \avg{ \partial_i \hat{E }_p\bkt{\mb{r}_0, t} \partial_i \hat{P }_k\bkt{\mb{r}_0, t- \tau}}}_\mr{(IIIB)}\right.\non\\
&\left.+ \underbrace{\avg{ \partial_i \hat{P }_p\bkt{\mb{r}_0, t} \partial_j \hat{E }_k\bkt{\mb{r}_0, t- \tau}}}_\mr{(IVA)}\underbrace{\avg{ \hat{E }_p\bkt{\mb{r}_0, t} \hat{P }_k\bkt{\mb{r}_0, t- \tau}}}_\mr{(IVB)} \right.\non\\
&\left.-\underbrace{ \avg{\partial_j \hat{P }_p\bkt{\mb{r}_0, t- \tau} \partial_i \hat{P }_k\bkt{\mb{r}_0, t}} }_\mr{(VA)}\underbrace{\avg{ \hat{E }_p\bkt{\mb{r}_0, t- \tau}  \hat{E }_k\bkt{\mb{r}_0, t}}}_\mr{(VB)}\right.\non\\
&\left.- \underbrace{\avg{ \hat{P }_p\bkt{\mb{r}_0, t- \tau} \hat{P }_k\bkt{\mb{r}_0, t}}}_\mr{(VIA)}\underbrace{ \avg{ \partial_j\hat{E }_p\bkt{\mb{r}_0, t- \tau} \partial_i \hat{E }_k\bkt{\mb{r}_0, t}}}_\mr{(VIB)}\right.\non\\
&\left.- \underbrace{\avg{ \hat{P }_p\bkt{\mb{r}_0, t- \tau} \hat{E }_k\bkt{\mb{r}_0, t}}}_\mr{(VIIA)}\underbrace{ \avg{ \partial_j \hat{E }_p\bkt{\mb{r}_0, t- \tau} \partial_i \hat{P }_k\bkt{\mb{r}_0, t}}}_\mr{(VIIB)}\right.\non\\
&\left.- \underbrace{\avg{ \partial_j \hat{P }_p\bkt{\mb{r}_0, t- \tau} \partial_i \hat{E }_k\bkt{\mb{r}_0, t}}}_\mr{(VIIIA)}\underbrace{\avg{ \hat{E }_p\bkt{\mb{r}_0, t- \tau} \hat{P }_k\bkt{\mb{r}_0, t}}  }_\mr{(VIIIB)} } ,
}
Substituting Eqs.~\eqref{eq:1a}--\eqref{eq:8b} in the above, we obtain the dissipation kernel as:
\eqn{
&\mc{D}_{ij} \bkt{\tau } = \frac{\hbar^2 \mu_0 ^2 }{2\pi ^2} \sum_{p,k}\int_0 ^\infty  \dd\omega \omega^2  \int _0 ^\infty \dd\omega' {\omega'} ^2\non\\
&\sbkt{ \im \cbkt{\partial_i\bkt{\dbar G\bkt{\mb{r}_0 ,\mb{r}_0 , \omega }}_{pk}\partial_j}\im \cbkt{ \bkt{\dbar G\bkt{\mb{r}_0 ,\mb{r}_0 , \omega' }}_{pk}} +\im \cbkt{\bkt{\dbar G\bkt{\mb{r}_0 ,\mb{r}_0 , \omega }}_{pk}}\im \cbkt{\partial_j \bkt{\dbar G\bkt{\mb{r}_0 ,\mb{r}_0 , \omega' }}_{pk}\partial_i }}\non\\
& \sbkt{\abs{\alpha\bkt{\omega}}^2\cbkt{\bkt{\bkt{n_\mr{th}\bkt{\omega}  +1}\bkt{n_\mr{th}\bkt{\omega'}  + 1} - n_\mr{th}\bkt{\omega}n_\mr{th}\bkt{\omega'}  } \sin \bkt{(\omega + \omega') \tau}   \right.\right.\non\\
& \left. \left.+ \bkt{\bkt{n_\mr{th}\bkt{\omega}  + 1}n_\mr{th}\bkt{\omega'}   - n_\mr{th}\bkt{\omega}\bkt{n_\mr{th}\bkt{\omega'}  +1}} \sin\bkt{(\omega - \omega') \tau}  }\right.\non\\
&\left.+  \cbkt{\bkt{\alpha\bkt{\omega}\alpha^\ast\bkt{\omega'}\bkt{n_\mr{th} \bkt{\omega} + 1}\bkt{n_\mr{th} \bkt{\omega'} + 1}-\alpha^\ast\bkt{\omega}\alpha\bkt{\omega'}n_\mr{th} \bkt{\omega}n_\mr{th} \bkt{\omega'}}}\sin\bkt{(\omega + \omega' )\tau} \right.\non\\
&\left.+ \cbkt{\bkt{\alpha (\omega)\alpha\bkt{\omega'}\bkt{n_\mr{th} (\omega) + 1} n_\mr{th}\bkt{\omega'} - \alpha ^\ast(\omega)\alpha^\ast\bkt{\omega'}n_\mr{th} (\omega) \bkt{n_\mr{th}\bkt{\omega'}  +1}} \sin \bkt{(\omega - \omega')\tau}}}.
}
The dissipation coefficient $ \Gamma_{ij} $ can be thus calculated using Eq.~\eqref{Eq:gammaij}.
 Noting that $ \int _ 0^\infty \dd\tau \tau \sin \bkt{(\omega - \omega ' )\tau } = -\pi  \pard{}{\omega} \delta\bkt{\omega - \omega'} $ and using integration by parts, we get  Eq.~\eqref{Eq:gammai}. 
 Similar to the decoherence coefficient, we can split up the Green tensor into its free and scattering parts to give:
 
\eqn{
\Gamma^\mr{free}_{ij}  =\frac{\hbar^2  \mu_0 ^2}{M\pi k_B T } \int_0 ^\infty  \dd\omega \omega^4 \abs{\alpha\bkt{\omega}}^2 n_\mr{th}\bkt{\omega}\bkt{n_\mr{th}\bkt{\omega} + 1} \sum_{p,k=\cbkt{x,y,z}}\im\sbkt{ \bkt{\dbar G_\mr{0} \bkt{\mb{r}_0 , \mb{r}_0 , \omega}} _{pk}} \im \sbkt{\partial_i\bkt{\dbar G_\mr{0} \bkt{\mb{r}_0 , \mb{r}_0 , \omega}}_{pk}\partial_j}}
\eqn{
\Gamma^\mr{s}_{ij}  =&\frac{\hbar^2  \mu_0 ^2}{M\pi k_B T } \int_0 ^\infty  \dd\omega \omega^4 \abs{\alpha\bkt{\omega}}^2 n_\mr{th}\bkt{\omega}\bkt{n_\mr{th}\bkt{\omega} + 1} \sum_{p,k=\cbkt{x,y,z}}\im\sbkt{ \bkt{\dbar G_\mr{0} \bkt{\mb{r}_0 , \mb{r}_0 , \omega}} _{pk}} \im \sbkt{\partial_i\bkt{\dbar G_\mr{s} \bkt{\mb{r}_0 , \mb{r}_0 , \omega}}_{pk}\partial_j}\non \\
+&\frac{\hbar^2  \mu_0 ^2}{M\pi k_B T } \int_0 ^\infty  \dd\omega \omega^4 \abs{\alpha\bkt{\omega}}^2 n_\mr{th}\bkt{\omega}\bkt{n_\mr{th}\bkt{\omega} + 1} \sum_{p,k=\cbkt{x,y,z}}\im\sbkt{ \bkt{\dbar G_\mr{s} \bkt{\mb{r}_0 , \mb{r}_0 , \omega}} _{pk}} \im \sbkt{\partial_i\bkt{\dbar G_\mr{0} \bkt{\mb{r}_0 , \mb{r}_0 , \omega}}_{pk}\partial_j}\non \\
+&\frac{\hbar^2  \mu_0 ^2}{M\pi k_B T } \int_0 ^\infty  \dd\omega \omega^4 \abs{\alpha\bkt{\omega}}^2 n_\mr{th}\bkt{\omega}\bkt{n_\mr{th}\bkt{\omega} + 1} \sum_{p,k=\cbkt{x,y,z}}\im\sbkt{ \bkt{\dbar G_\mr{s} \bkt{\mb{r}_0 , \mb{r}_0 , \omega}} _{pk}} \im \sbkt{\partial_i\bkt{\dbar G_\mr{s} \bkt{\mb{r}_0 , \mb{r}_0 , \omega}}_{pk}\partial_j},
 } 
 where the interpretation of the terms follows the same line of reasoning as for the decoherence coefficient. Note that each line in the above expression contains distinct combinations of free-space ( $ \dbar G_0 \bkt{\mb{r}_0, \mb{r}_0, \omega}$) and scattered Green tensors $ \dbar G_s \bkt{\mb{r}_0, \mb{r}_0, \omega}$.

\section{Momentum fluctuations}
\label{App:MomFl}
In the following the indices $i,j$ will take the values $i,j=\cbkt{x,y}$ and the indices $p,k$ will take the values $p,k=\cbkt{x,y,z}$
\eqn{
\avg{\Delta p_i\Delta p_j} = \frac{1}{4}\int_0 ^{\Delta t}\dd\tau_1 \int_0 ^{\Delta t}\dd\tau_2&\avg{  \sbkt{\partial_i \hmb{P }\bkt{\mb{r}_0, \tau_1}\cdot \hmb{E }\bkt{\mb{r}_0,\tau_1} + \hmb{P }\bkt{\mb{r}_0, \tau_1}\cdot \partial_i \hmb{E }\bkt{\mb{r}_0, \tau_1}}\right.\non\\
&\left.\sbkt{\partial_j \hmb{P }\bkt{\mb{r}_0,\tau_2}\cdot \hmb{E }\bkt{\mb{r}_0, \tau_2} + \hmb{P }\bkt{\mb{r}_0, \tau_2}\cdot \partial_j \hmb{E }\bkt{\mb{r}_0, \tau_2}}}
}
Using the statistical independence of $\cbkt{ \hat E_j , \hat P_j} $ and their spatial derivatives $\cbkt{ \partial_i \hat E_j , \partial_i \hat P_j} $:
\eqn{
\avg{\Delta p_i\Delta p_j}=& \underbrace{ \frac{1}{4}\int_0 ^{\Delta t}\dd\tau_1 \int_0 ^{\Delta t}\dd\tau_2\sum_{p,k}\avg{\partial_i \hat{P }_p\bkt{\mb{r}_0, \tau_1} \partial_j \hat{P }_k\bkt{\mb{r}_0, \tau_2}}\avg{ \hat{E }_p\bkt{\mb{r}_0, \tau_1}  \hat{E }_k\bkt{\mb{r}_0, \tau_2}}}_\mr{(I)}\non\\
&+\underbrace{\frac{1}{4}\int_0 ^{\Delta t}\dd\tau_1 \int_0 ^{\Delta t}\dd\tau_2\sum_{p,k}\avg{ \hat{P }_p\bkt{\mb{r}_0, \tau_1} \hat{P }_k\bkt{\mb{r}_0, \tau_2}}\avg{ \partial_i\hat{E }_p\bkt{\mb{r}_0, \tau_1} \partial_j \hat{E }_k\bkt{\mb{r}_0, \tau_2}}}_\mr{(II)}\non\\
&+\underbrace{\frac{1}{4}\int_0 ^{\Delta t}\dd\tau_1 \int_0 ^{\Delta t}\dd\tau_2\sum_{p,k} \avg{ \hat{P }_p\bkt{\mb{r}_0, \tau_1} \hat{E }_k\bkt{\mb{r}_0, \tau_2}} \avg{ \partial_i \hat{E }_p\bkt{\mb{r}_0, \tau_1} \partial_j \hat{P }_k\bkt{\mb{r}_0, \tau_2}}}_\mr{(III)}\non\\
&+\underbrace{\frac{1}{4}\int_0 ^{\Delta t}\dd\tau_1 \int_0 ^{\Delta t}\dd\tau_2\sum_{p,k} \avg{ \partial_i \hat{P }_p\bkt{\mb{r}_0, \tau_1} \partial_j \hat{E }_k\bkt{\mb{r}_0, \tau_2}}\avg{ \hat{E }_p\bkt{\mb{r}_0, \tau_1} \hat{P }_k\bkt{\mb{r}_0,  \tau_2}} }_\mr{(IV)}
}

Let us consider the first term in the above equation more carefully:
\eqn{
\mr{(I)}= &\frac{\hbar^2\mu_0 ^2}{4 \pi ^2}\int_0 ^\infty \dd\omega \omega^2\int_0 ^\infty \dd\omega' {\omega'}^2 \int_0 ^{\Delta t}\dd\tau_1 \int_0 ^{\Delta t}\dd\tau_2\sum_{p,k}\abs{\alpha\bkt{\omega}}^2\sum_{p,k}\im\sbkt{ \bkt{\dbar G \bkt{\mb{r}_0 , \mb{r}_0 , \omega'}} _{pk}} \im \sbkt{\partial_i\bkt{\dbar G \bkt{\mb{r}_0 , \mb{r}_0 , \omega}}_{pk}\partial_j}\non\\
& \sbkt{\bkt{n_\mr{th}\bkt{\omega} + 1} e^{-i \omega \bkt{\tau_1 - \tau_2} }+ n_\mr{th}\bkt{\omega}  e^{i \omega \bkt{\tau_1 - \tau_2}}}\sbkt{\bkt{n_\mr{th}\bkt{\omega'} + 1} e^{-i \omega '\bkt{\tau_1 - \tau_2} }+ n_\mr{th}\bkt{\omega'}  e^{i \omega'\bkt{\tau_1 - \tau_2} }}\\
=&\frac{\hbar^2\mu_0 ^2}{4 \pi ^2}\int_0 ^\infty \dd\omega \omega^2\int_0 ^\infty \dd\omega' {\omega'}^2 \int_0 ^{\Delta t}\dd\tau_1 \int_0 ^{\Delta t}\dd\tau_2\sum_{j,k}\abs{\alpha\bkt{\omega}}^2\sum_{p,k}\im\sbkt{ \bkt{\dbar G \bkt{\mb{r}_0 , \mb{r}_0 , \omega'}} _{pk}} \im \sbkt{\partial_i\bkt{\dbar G \bkt{\mb{r}_0 , \mb{r}_0 , \omega}}_{pk}\partial_j}\non\\
& \sbkt{\bkt{n_\mr{th}\bkt{\omega} + 1}\bkt{n_\mr{th}\bkt{\omega'} + 1} e^{-i \bkt{\omega + \omega'}\bkt{\tau_1 - \tau_2} }+ n_\mr{th}\bkt{\omega}  n_\mr{th}\bkt{\omega'}  e^{i\bkt{ \omega+ \omega'} \bkt{\tau_1 - \tau_2}} \right.\non\\
&\left.+ \bkt{n_\mr{th}\bkt{\omega}  +1}n_\mr{th}\bkt{\omega'}  e^{-i\bkt{ \omega- \omega'} \bkt{\tau_1 - \tau_2}}+ n_\mr{th}\bkt{\omega} \bkt{n_\mr{th}\bkt{\omega'} + 1}  e^{i\bkt{ \omega- \omega'} \bkt{\tau_1 - \tau_2}}}\\
=&\frac{\hbar^2\mu_0 ^2}{ \pi ^2}\int_0 ^\infty \dd\omega \omega^2\int_0 ^\infty \dd\omega' {\omega'}^2 \sum_{p,k}\abs{\alpha\bkt{\omega}}^2\sum_{p,k}\im\sbkt{ \bkt{\dbar G \bkt{\mb{r}_0 , \mb{r}_0 , \omega'}} _{pk}} \im \sbkt{\partial_i\bkt{\dbar G \bkt{\mb{r}_0 , \mb{r}_0 , \omega}}_{pk}\partial_j}\non\\
& \sbkt{\cbkt{\bkt{n_\mr{th}\bkt{\omega} + 1}\bkt{n_\mr{th}\bkt{\omega'} + 1}+ n_\mr{th}\bkt{\omega}  n_\mr{th}\bkt{\omega'}}\frac{\sin^2\bkt{\bkt{\omega + \omega ' }\Delta t/2}}{\bkt{\omega + \omega'}^2}\right.\non\\
&\left.+ \cbkt{\bkt{n_\mr{th}\bkt{\omega}  +1}n_\mr{th}\bkt{\omega'} + n_\mr{th}\bkt{\omega} \bkt{n_\mr{th}\bkt{\omega'} + 1}} \frac{\sin^2 \bkt{(\omega - \omega ')\Delta t/2}}{\bkt{\omega - \omega'}^2} }
} 
We note that in the limit where  $ \Delta t $ is  small, the term proportional to $ \sim \frac{\sin^2\bkt{\bkt{\omega + \omega ' }\Delta t/2}}{\bkt{\omega + \omega'}^2}$ vanishes, and  $  \int_0 ^\infty \dd\omega' \frac{\sin^2\bkt{\bkt{\omega - \omega ' }\Delta t/2}}{\bkt{\omega - \omega'}^2} \approx \frac{\pi \Delta t}{2} $.
This yields:
\eqn{\mr{(I)} = \frac{\hbar^2\mu_0 ^2\Delta t}{ \pi} \int_0 ^\infty \dd\omega \omega^4  n_\mr{th}\bkt{\omega}\bkt{n_\mr{th}\bkt{\omega} + 1}\abs{\alpha \bkt{\omega}}^2\sum_{p,k}\im\sbkt{ \bkt{\dbar G \bkt{\mb{r}_0 , \mb{r}_0 , \omega}} _{pk}} \im \sbkt{\partial_i\bkt{\dbar G \bkt{\mb{r}_0 , \mb{r}_0 , \omega}}_{pk}\partial_j}.
}

Similarly, we can simplify the remaining terms to find $ \mr{(I)} = \mr{(II)} = \mr{(III)} = \mr{(IV)} $, which gives us the momentum diffusion constant:

\eqn{
D_{ij}= \frac{4\hbar^2\mu_0 ^2}{\pi} \int_0 ^\infty \dd\omega \omega^4  n_\mr{th}\bkt{\omega}\bkt{n_\mr{th}\bkt{\omega} + 1}\abs{\alpha \bkt{\omega}}^2\sum_{p,k=\cbkt{x,y,z}}\im\sbkt{ \bkt{\dbar G \bkt{\mb{r}_0 , \mb{r}_0 , \omega}} _{pk}} \im \sbkt{\partial_i\bkt{\dbar G \bkt{\mb{r}_0 , \mb{r}_0 , \omega}}_{pk}\partial_j},
}

Which we can once again split up into free and scattering parts: 

\eqn{
D^\mr{free}_{ij}= &\frac{4\hbar^2\mu_0 ^2}{\pi} \int_0 ^\infty \dd\omega \omega^4  n_\mr{th}\bkt{\omega}\bkt{n_\mr{th}\bkt{\omega} + 1}\abs{\alpha \bkt{\omega}}^2\sum_{p,k=\cbkt{x,y,z}}\im\sbkt{ \bkt{\dbar G_\mr{0} \bkt{\mb{r}_0 , \mb{r}_0 , \omega}} _{pk}} \im \sbkt{\partial_i\bkt{\dbar G_\mr{0}  \bkt{\mb{r}_0 , \mb{r}_0 , \omega}}_{pk}\partial_j}}

\eqn{
D^\mr{s}_{ij}= &\frac{4\hbar^2\mu_0 ^2}{\pi} \int_0 ^\infty \dd\omega \omega^4  n_\mr{th}\bkt{\omega}\bkt{n_\mr{th}\bkt{\omega} + 1}\abs{\alpha \bkt{\omega}}^2\sum_{p,k=\cbkt{x,y,z}}\im\sbkt{ \bkt{\dbar G_\mr{0}  \bkt{\mb{r}_0 , \mb{r}_0 , \omega}} _{pk}} \im \sbkt{\partial_i\bkt{\dbar G_\mr{s}  \bkt{\mb{r}_0 , \mb{r}_0 , \omega}}_{pk}\partial_j}\non\\
+&\frac{4\hbar^2\mu_0 ^2}{\pi} \int_0 ^\infty \dd\omega \omega^4  n_\mr{th}\bkt{\omega}\bkt{n_\mr{th}\bkt{\omega} + 1}\abs{\alpha \bkt{\omega}}^2\sum_{p,k=\cbkt{x,y,z}}\im\sbkt{ \bkt{\dbar G_\mr{s}  \bkt{\mb{r}_0 , \mb{r}_0 , \omega}} _{pk}} \im \sbkt{\partial_i\bkt{\dbar G_\mr{0}  \bkt{\mb{r}_0 , \mb{r}_0 , \omega}}_{pk}\partial_j}\non\\
+&\frac{4\hbar^2\mu_0 ^2}{\pi} \int_0 ^\infty \dd\omega \omega^4  n_\mr{th}\bkt{\omega}\bkt{n_\mr{th}\bkt{\omega} + 1}\abs{\alpha \bkt{\omega}}^2\sum_{p,k=\cbkt{x,y,z}}\im\sbkt{ \bkt{\dbar G_\mr{s}  \bkt{\mb{r}_0 , \mb{r}_0 , \omega}} _{pk}} \im \sbkt{\partial_i\bkt{\dbar G_\mr{s}  \bkt{\mb{r}_0 , \mb{r}_0 , \omega}}_{pk}\partial_j}.
}


\end{widetext}

\bibliography{BM}
\end{document}